\documentclass[times,astrosymb,twocolumn]{aastex631}
\usepackage{xspace}
\usepackage[utf8]{inputenc}

\newcommand{\paa}{Pa\ensuremath{\alpha}}

\newcommand{\methanol}{\ensuremath{\textrm{CH}_3\textrm{OH}}\xspace}

\newcommand{\water}{H$_{2}$O\xspace}		

\newcommand{\kms}{\textrm{km~s}\ensuremath{^{-1}}\xspace}	

\newcommand{\percc}{\ensuremath{\textrm{cm}^{-3}}\xspace}

\newcommand{\persc}{\ensuremath{\textrm{cm}^{-2}}\xspace}

\newcommand{\um}{\ensuremath{\mu \textrm{m}}\xspace}    

\def\ee#1{\ensuremath{\times10^{#1}}}

\def\eqref#1{Equation \ref{#1}}

\usepackage{tikz}
\usetikzlibrary{calc,tikzmark} 

\newcommand\nstarstotal{377,236\xspace}
\newcommand\nstarsallgood{56,146\xspace}
\newcommand\nstarsccd{34,150\xspace}
\newcommand\bra{Br\ensuremath{\alpha}\xspace}
\newcommand\pfb{Pf\ensuremath{\beta}\xspace}
\newcommand\hue{Hu\ensuremath{\epsilon}\xspace}
\def\rr#1{#1}
\def\fix#1{\textbf{#1}}

\begin{document}
\title{JWST reveals widespread CO ice and gas absorption in the Galactic Center cloud G0.253+0.016}
\author[0000-0001-6431-9633]{Adam Ginsburg}
\affiliation{Department of Astronomy, University of Florida, P.O. Box 112055, Gainesville, FL, USA}
\author[0000-0003-0410-4504]{Ashley T. Barnes}
\affiliation{European Southern Observatory (ESO), Karl-Schwarzschild-Stra{\ss}e 2, 85748 Garching, Germany}
\author[0000-0002-6073-9320]{Cara D. Battersby}
\affiliation{University of Connecticut, Department of Physics, 196A Auditorium Road, Unit 3046,
Storrs, CT 06269}
\author[0000-0002-4407-885X]{Alyssa Bulatek}
\affiliation{Department of Astronomy, University of Florida, P.O. Box 112055, Gainesville, FL, USA}
\author[0000-0002-1313-429X]{Savannah Gramze}
\affiliation{Department of Astronomy, University of Florida, P.O. Box 112055, Gainesville, FL, USA}
\author[0000-0001-9656-7682]{Jonathan D. Henshaw}
\affiliation{Astrophysics Research Institute, Liverpool John Moores University, 146 Brownlow Hill, Liverpool L3 5RF, UK}
\affiliation{Max-Planck-Institut f\"ur Astronomie, K\"onigstuhl 17, D-69117 Heidelberg, Germany}
\author[0000-0003-0416-4830]{Desmond Jeff}
\affiliation{Department of Astronomy, University of Florida, P.O. Box 112055, Gainesville, FL, USA}
\author[0000-0003-2619-9305]{Xing Lu}
\affiliation{Shanghai Astronomical Observatory, Chinese Academy of Sciences, 80 Nandan Road, Shanghai 200030, P.\ R.\ China}
\author[0000-0001-8782-1992]{E. A. C. Mills}
\affiliation{Department of Physics and Astronomy, University of Kansas, 1251 Wescoe Hall Dr., Lawrence, KS 66045, USA}
\author[0000-0001-7330-8856]{Daniel L. Walker}
\affiliation{UK ALMA Regional Centre Node, Jodrell Bank Centre for Astrophysics, The University of Manchester, Manchester M13 9PL, UK}

\date{September 2023}

\begin{abstract}
We report JWST NIRCam observations of G0.253+0.016, the molecular cloud in the Central Molecular Zone known as ``The Brick,'' with the F182M, F187N, F212N, F410M, F405N, and F466N filters.
We catalog \nstarsallgood stars detected in all six filters using the \texttt{crowdsource} package.
Stars within and behind The Brick exhibit prodigious absorption in the F466N filter that is produced by a combination of CO ice and gas.
In support of this conclusion, and as a general resource, we present models of CO gas and ice and CO$_2$ ice in the F466N, F470N, and F410M filters.
Both CO gas and ice contribute to the observed stellar colors.
We show, however, that CO gas does not absorb the \pfb and \hue lines in F466N, but that these lines show excess absorption, indicating that CO ice is present and contributes to observed F466N absorption.
The most strongly absorbed stars in F466N are extincted by $\sim2$ magnitudes, corresponding to $>80\%$ flux loss.
This high observed absorption requires very high column densities of CO, and thus a total CO column that is in tension with standard CO abundance and/or gas-to-dust ratios.
\rr{This result suggests the CO/H$_2$ ratio and dust-to-gas ratio are greater} in the Galactic Center than the Galactic disk.
Ice and/or gas absorption is observed even in the cloud outskirts, implying that additional caution is needed when interpreting stellar photometry in filters that overlap with ice bands throughout galactic centers.
\end{abstract}

\section{Introduction}

G0.253+0.016, AKA ``The Brick'', is among the best-studied infrared dark clouds in the Galaxy \citep{Lis1991,Lis1994a,Lis1994b,Lis1998,Lis2001,Longmore2012,Kauffmann2013,Rodriguez2013,Clark2013,Rathborne2014a,Rathborne2014b,Rathborne2015,Johnston2014,Bally2014,Pillai2015,Federrath2016,Marsh2016,Walker2016,Henshaw2019,Henshaw2022,Petkova2023}.
It is well-known for being dense and turbulent \citep{Clark2013,Rathborne2015,Federrath2016,Mills2018,Henshaw2019,Henshaw2020} while exhibiting few signs of star formation, much less than is typical for such a massive cloud \citep{Longmore2012,Rodriguez2013,Mills2015,Walker2016,Walker2021}.
Several explanations have been offered for its relatively poor star formation: that it is young \citep{Kruijssen2015,Henshaw2016}, that it is highly turbulent \citep{Federrath2016}, that it is supported by magnetic fields \citep{Pillai2015}, and that it is many clouds along the line of sight \citep{Henshaw2019,Henshaw2022}.
Each of these explanations is likely to play some role in the cloud's state and evolution.

Gas in the Galactic Center is notably different from gas seen elsewhere in our Galaxy.
It is richer in complex molecules \citep{Jones2012} and warmer \citep{Ao2013,Ginsburg2016,Krieger2017}.
Despite high gas temperatures, the dust in the CMZ is not terribly warm \citep{Tang2021}, so ice can accumulate on dust grains.
Ice has long been seen in Galactic Center mid-infrared spectra \citep{Lutz1996,Chiar2000,Moneti2001}, as has CO gas. 
Toward Sgr A*, ice comprises a minority ($<10\%$) of the CO, which is dominated instead by gas, as expected given the high gas temperatures and extreme velocity dispersion of gas in the inner parsec \citep{Moneti2001,Moultaka2009}.
Beyond the inner parsec, though, the ice properties of the Galactic Center are little explored.
A handful of Infrared Space Observatory (ISO) spectra were taken toward various positions, revealing both CO$_2$ and CO ice features, but little has been written about them.
\citet{An2011} and \citet{Jang2022} used Spitzer IRS spectra to show that CO$_2$ ice is common toward massive young stellar objects (MYSOs) in the GC, and that CO$_2$ is present in both gas and ice phases.

\rr{More generally, ice is observed throughout the molecular interstellar medium \citep{Boogert2015}.
However, ice is much more poorly studied than gas in the interstellar medium because observable ice features occur only in the infrared, either in narrow, difficult-to-observe bands from the ground \citep[e.g.][]{Gunay2020,Gunay2022}, or in bands observable only from from space.
The majority of published ice studies use spectroscopy, not photometry, in large part because the effects of ice on broadband filters (e.g., Spitzer's IRAC) are usually small.
The wide range of medium- and narrow-band filters on JWST NIRCam change the state of the field, enabling extensive broad-field ice study through photometry.
}

We present JWST observations in narrow-band filters toward The Brick, highlighting the first striking result that CO ice is widespread.
In Section \ref{sec:dataprocessing}, we describe the data processing and catalog creation.
Section \ref{sec:results} describes the measurements of both star colors (\S \ref{sec:bluestars}) and diffuse gas emission (\S \ref{sec:recombination}), then describes models of both CO gas (\S \ref{sec:cogas}) and ice (\S \ref{sec:coice}) absorption that explain some of the observed colors.
We briefly discuss these results in Section \ref{sec:icegasdiscussion} and then conclude in Section \ref{sec:conclusion}.
All of the analysis tools, including the notebooks used to make the figures in this document, are made available through a GitHub repository\footnote{\url{https://github.com/keflavich/brick-jwst-2221/};  DOI: 10.5281/zenodo.8313307}.

\section{Observations}
Observations were taken on \rr{August 28, 2022} as part of JWST program 2221 in visit 001.
\rr{The data presented in this paper were obtained from the Mikulski Archive for Space Telescopes (MAST) at the Space Telescope Science Institute (STScI). The specific observations analyzed can be accessed via \dataset[DOI 10.17909/2ffq-e139]{https://doi.org/10.17909/2ffq-e139}.}
This program consists of two observations focused on The Brick, with coordinated parallel observations performed toward \rr{Central Molecular Zone} Cloud C.
We present only the Brick NIRCam observations in this work; the MIRI observations of The Brick and NIRCam and MIRI observations of Cloud C will be presented in future works.
We obtained images in \rr{six} filters listed in Table \ref{tab:observations}.
We observed in narrow-band filters in order to measure the extended line emission from hydrogen recombination lines (\paa, \bra, \pfb) and to search for outflows (H$_2$ in F212N) and hot CO emission from disks (F466N).
The GTO program 1182 has observed approximately the same field in broad-band filters.

\begin{figure*}[htp]
    \centering
    \includegraphics[width=\textwidth]{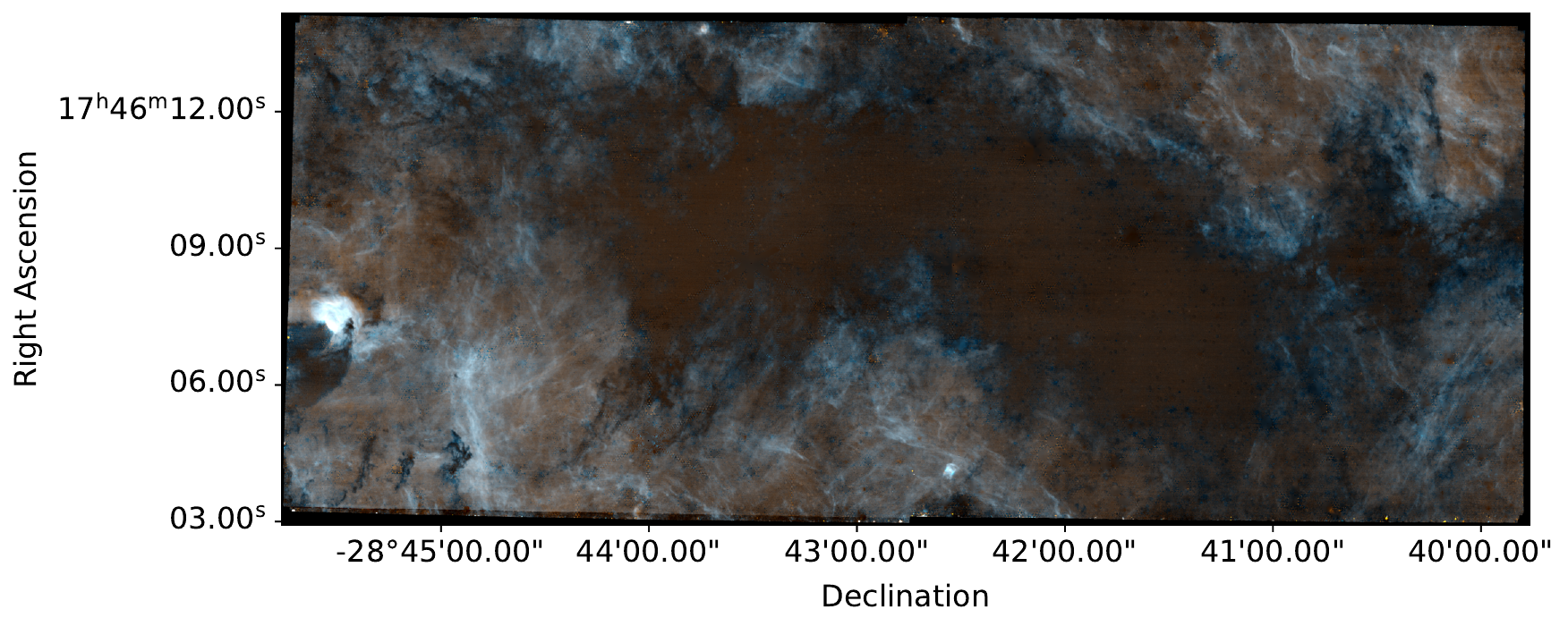}
    \caption{The full-field star-subtracted image as described in \S \ref{sec:starsub}.
    This image captures extended recombination line emission from Pf$\beta$ and Hu$\epsilon$ in F466N
    \rr{(466m410, continuum-subtracted F466N, is shown in red) and Br$\alpha$ in F405N (405m410. continuum-subtracted F405N, is shown in blue), highlighting The Brick as the dark cloud spanning the image.}
    \rr{The green channel is the sum of the red and blue channels and therefore adds no information.}
    Most of the \rr{extended emission is approximately white} because both the F466N and F405N filters contain recombination lines (Pf$\beta$+\hue and Br$\alpha$, respectively).
    The color is \rr{more red} when dust extinction is the dominant absorption effect and \rr{more blue} when CO absorption is dominant (see \S \ref{sec:recombination})\rr{, but the colors should not be interpreted quantitatively}.
    Note that significant artistic license has been taken to fill in regions where star PSFs have made pixel fluxes unrecoverable; the image represents the overall structure of the cloud well, but many small-scale features are artifacts of the star subtraction.
    Declination is on the X-axis, contrary to convention.
    A full resolution version is available online
    (which we hope to make available through the journal; for now use \url{https://www.dropbox.com/scl/fi/39cfq8yr40460qaoy1wlf/BrickJWST_merged_longwave_narrowband_lighter.png?rlkey=lusbv81fsr9rvupt99zqmvpqj&dl=0}).
    }
    \label{fig:fullfieldstarless}
\end{figure*}

In the NIRCam data we present, each image is comprised of 24 exposures taken in the 6-TIGHT FULLBOX mosaic strategy,\footnote{\url{https://jwst-docs.stsci.edu/jwst-near-infrared-camera/nircam-operations/nircam-dithers-and-mosaics/nircam-primary-dithers}} with \rr{six} independent positions and \rr{four} subpixel dithers per position.
Frames were read out in BRIGHT-2 mode with 2 groups per integration for a total exposure time of 1031 seconds.\footnote{\url{https://jwst-docs.stsci.edu/jwst-near-infrared-camera/nircam-instrumentation/nircam-detector-overview/nircam-detector-readout-patterns}.}

\section{Data processing}
\label{sec:dataprocessing}

We downloaded the data from the MAST archive using \texttt{astroquery} \citep{Ginsburg2019,Brasseur2020}.
We reprocessed data starting from L2 products, 
i.e., the \texttt{cal} files, which include 24 individual flux-calibrated frames for each filter.

Our reduction code is provided on GitHub\footnote{\url{https://github.com/keflavich/brick-jwst-2221/releases/tag/resubmission_20230903};  DOI: 10.5281/zenodo.8313307}.

\subsection{Frame matching astrometry}
Images were processed with a slightly modified version of the JWST pipeline \rr{based on version 1.11.1 \citep{bushouse_howard_2023_8099867}}.
The \texttt{tweakreg} command was run on long-wavelength NIRCam data using the VVV DR2 catalog \citep{Saito2012} as an astrometric reference instead of the Gaia catalog (there were too few stars detected in common \rr{by both} Gaia and JWST, and most were saturated).
We then created a reference catalog based on the F405N catalog, which had fewer saturated bright stars than F410M and therefore more good associations with the ground-based NIR data. 
In the \texttt{make\_reftable.py} script, we cut the F405N catalog based on \texttt{crowdsource} quality flags (\texttt{qf}$>0.95$, \texttt{spread} $<0.25$, \texttt{frac} $>0.9$).
This reference catalog was then used as the input to \texttt{tweakreg} for the other filters.
We found that the \rr{\texttt{tweakreg}} pipeline did not adequately correct the image registration to the absolute coordinates we provided, so we manually cross-matched the catalogs and computed shifts using the \texttt{realign\_to\_catalog} and \texttt{merge\_a\_to\_b} functions in \texttt{align\_to\_catalogs.py}.

\subsection{1/f noise removal}
\label{sec:oneoverf}

The narrow-band filters, particularly F187N, F212N, and F466N, exhibited significant `streaking' noise that is strongly evident in the low-signal regions of the image, i.e., the majority of the molecular cloud.
This streaking \rr{was} caused by 1/f noise in the detectors (STScI helpdesk ticket INC0181624).
As a first pre-processing step, we performed `destreaking' on each detector following a method suggested by Massimo Robberto (priv. comm.), in which:
\begin{enumerate}
    \item Each detector \rr{was} split into four horizontal quadrants with width 512 pixels and height 2048 pixels.
    \item The median across the horizontal axis \rr{was} calculated, resulting in a 2048-pixel array.
    \item In a slight departure from Massimo's method, we then smooth\rr{ed} the median array using a 1D median filter with length that varie\rr{d} depending on the filter.
    \texttt{{'F410M': 15, 'F405N': 256, 'F466N': 55,  'F182M': 55, 'F187N': 256, 'F212N': 512}}.  We found that the original method, which was to obtain a single constant at this step, turned the destreaking process into a high-pass filter and therefore removed significant extended emission.
    \item We subtract\rr{ed} the median array from each quadrant, then add\rr{ed} back the smoothed median.
\end{enumerate}

We evaluated the effectiveness of this process by eye.
While the original destreaker completely removed the 1/f horizontal features, it also removed all of the extended background.
The modified version removed most of the horizontal features while preserving the large-scale extended structure.
It is likely that some intermediate-scale features (i.e., physical features comparable to 512 pixels across) are not recovered by this procedure, which will need to be accounted for in analysis of the extended emission.

\subsection{Photometric and Astrometic Cataloging}
\label{sec:photometry}
For photometry of unsaturated stars, we use\rr{d} the \texttt{crowdsource} python package \citep{Schlafly2021}.
We used a PSF model from \texttt{webbpsf} \citep{Perrin2015}.
Because we were using mosaiced images, the \texttt{webbpsf} PSF is not a perfect representation of the data; each individual frame had to be shifted and drizzled to form our final images.
Furthermore, for the short wavelength bands, we adopted the PSF for a single detector (NRCA1 or NRCB1 as appropriate) for the full frame, since \texttt{webbpsf} does not provide a tool to produce a PSF grid across the whole module.

\rr{We identified saturated stars so that we could ignore them, and stars too close to them (which are likely to be affected by the extended PSFs of saturated stars), for further analysis.}
\rr{To identify these stars, we measured the centroids of all regions in which either the data (FITS extension \texttt{SCI}) or the variance (FITS extension \texttt{VAR\_POISSON}) was zero.}
\rr{We then fitted the PSF of these stars excluding the saturated pixels and a surrounding set of pixels identified through binary dilation.}
\rr{The detailed values of the dilation size are given in \texttt{saturated\_star\_finding.py}.}
\rr{While we measured both photometry and astrometry of these saturated stars, we use only the astrometry in subsequent sections, and only as a means to automatically exclude saturated stars and their nearest neighbors.}

\subsection{Catalog matching}
\label{sec:catalogxmatch}
We assemble\rr{d} a catalog consisting of all sources found in any of our six filters.
To assemble the coordinate list, we start\rr{ed} with all coordinates in the F405N catalog, then for each other filter, we add\rr{ed} all sources that \rr{did} not have a match in the existing catalog within $d<0.15\arcsec$.
The crossmatch shows a large peak for matches within $d<0.1\arcsec$ with large tails at greater separation\rr{; we have not investigated the origin of these large-offset sources, but suspect low signal-to-noise sources, PSF artifacts, and features in the extended background may contribute}.
\rr{We then excluded all sources with a crossmatch distance to the reference filter's coordinates $d>0.1\arcsec$ (the reference filter is F405N by default, but for sources with nondetections in F405N, a different reference filter was adopted).}

For subsequent analysis, we then reject\rr{ed} all sources with magnitude errors $\sigma_m > 0.1$, `quality factor' \texttt{qf} $<=0.6$, \texttt{spread} $>0.25$, or \texttt{fracflux} $<0.8$\rr{, all of which are values calculated by \texttt{crowdsource}}.
These choices select for round, pointlike, unblended stars.
\rr{We also limit\rr{ed} our analysis to sources with detections in all six bands.}

The resulting crossmatched catalogs had RMS positional offsets $<0.02$ \arcsec (Table \ref{tab:observations}). 
We do not concern ourselves further with astrometry in this manuscript, but caution that\rr{, with these sizeable crossmatch errors,} our catalogs likely are not yet of sufficient quality to support proper motion measurements.

We \rr{found} \nstarstotal stars with a good measurement in at least one filter, and \nstarsallgood with good measurements in all \rr{six} filters.
The number \rr{of good measurements} found in each filter is given in Table \ref{tab:observations} (these include sources with offsets from the reference filter $d>0.1\arcsec$).

\begin{table}
\caption{Observations \label{tab:observations}}
\begin{tabular}{cccc}
\hline \hline
Filter Name & RMS Offset & 90th percentile & \# of sources \\
 & $\mathrm{{}^{\prime\prime}}$ & $\mathrm{mag}$$\mathrm{\left( \mathrm{AB} \right)}$ &  \\
\hline
F182M & 0.020 & 20.4 & 337561 \\
F187N & 0.021 & 20.3 & 213894 \\
F212N & 0.020 & 19.5 & 236077 \\
F405N & - & 19.5 & 85126 \\
F410M & 0.016 & 19.6 & 102344 \\
F466N & 0.021 & 19.6 & 79629 \\
\hline
\end{tabular}
\par
The RMS offset reports the standard deviation of the source position difference between the specified filter and the reference filter, F405N.
The 90th percentile column reports the 90th percentile magnitude in the catalog to give a general sense of depth.
\end{table}

\subsection{Starless Image Creation}
\label{sec:starsub}
For comparison of star locations to extinction features, we preferr\rr{ed} to work with an image with stars removed.
\rr{Note that, because of significant uncertainty in this process, we have used the star-subtracted images only for qualitative, not quantitative, analysis.}
A starless image is the natural residual of an image that has been processed through a PSF photometry routine that appropriately accounts for the non-point-source background.
However, \rr{when we produced} such images, \rr{they had} substantial residual features\rr{, which were caused by a combination of an imperfect PSF model and oversubtraction of sources on extended backgrounds}.
To create a cleaner starless image, we \rr{took} the difference between the narrow-band and medium-band images after appropriately scaling the narrow-band image.
The F405N image was convolved with a 0.3 pixel Gaussian to better match the PSF of the F410M filter.

\begin{figure*}
    \centering

    \begin{tikzpicture}
        \node[anchor=north west,inner sep=0pt] at (0,0){\includegraphics[width=\textwidth]{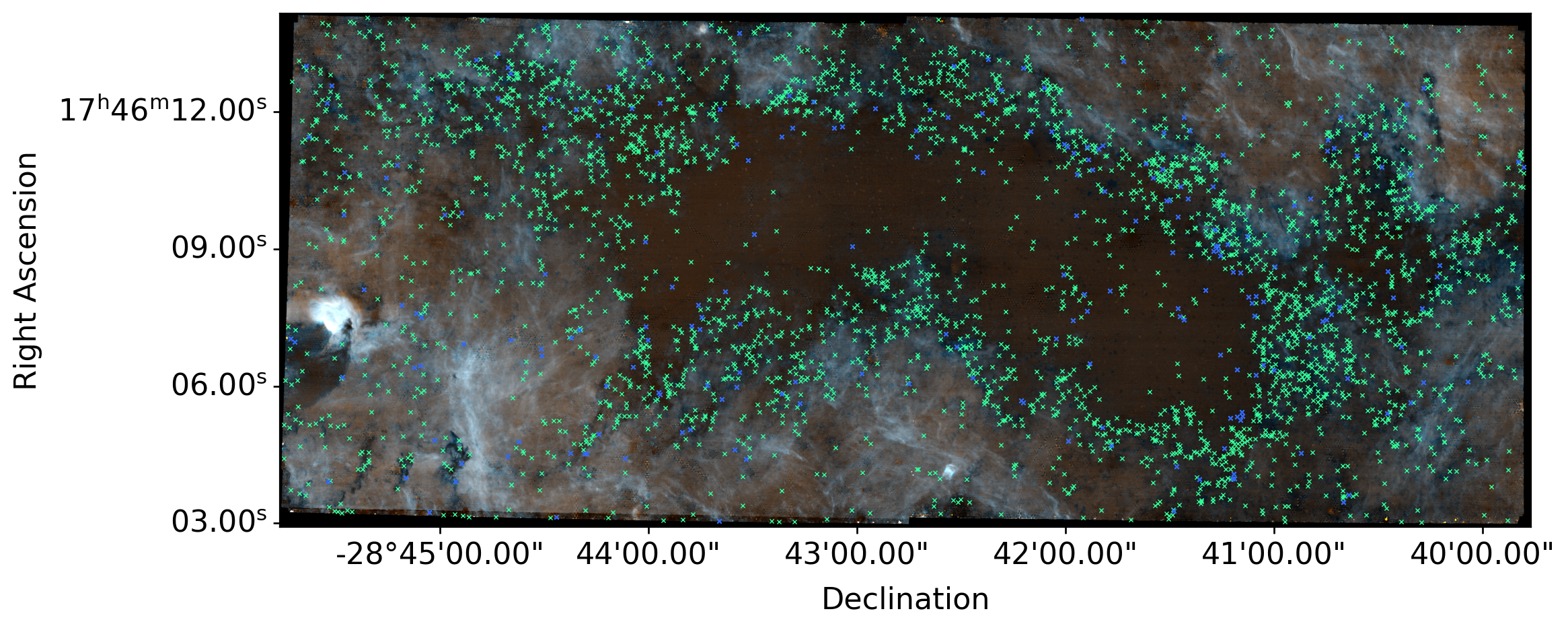}};
        \node[font=\bfseries\large] at (20ex,-3ex) {(a)};
    \end{tikzpicture}
    \begin{tikzpicture}
        \node[anchor=north west,inner sep=0pt] at (0,0){\includegraphics[width=\textwidth]{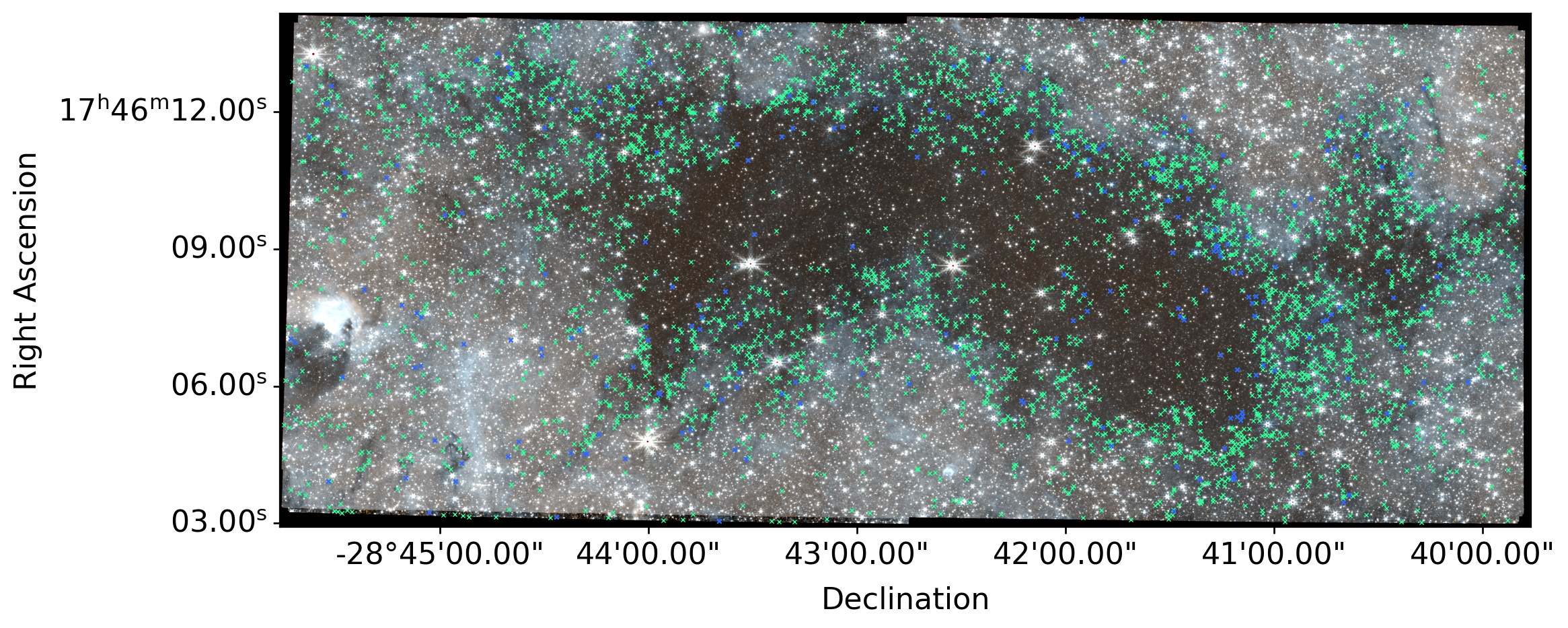}};
        \node[font=\bfseries\large] at (20ex,-3ex) {(b)};
    \end{tikzpicture}
    \caption{Stars with blue \rr{F410M-F466N} colors ([F410M]-[F466N] \fix{$< -0.45$} \rr{mag} in green and \fix{$<-1.45$} \rr{mag} in blue) shown with X's on the star-subtracted RGB  (\rr{a}, top) and not-star-subtracted (\rr{b}, bottom) image.
    Note that declination is on the X-axis, contrary to convention.
    }
    \label{fig:bluestarsextended}
    \label{fig:bluestarsonstars}
\end{figure*}

We produce\rr{d} a \textit{line-free} F410M image, labeled 410m405, \rr{by the following equation: 
\begin{equation}
S_{410m405} = \frac{S_{F410M} - S_{F405N} \cdot w}{1 - w}
\end{equation}
where $w$ is the fractional bandwidth of F410M covered by F405N \rr{and $S_{filtername}$ is the surface brightness in a given filter}.
This process effectively creates a continuum-only `notch' filter image}.
We produce\rr{d} a \textit{star-free} F405N image, labeled 405m410, by subtracting the (theoretically continuum-only) 410m405 image \rr{from F405N}.
\rr{Note that the images are in units of surface brightness, MJy sr$^{-1}$, such that line emission in broad-band filters is diluted (will have a lower surface brightness), while spectrally flat continuum sources (to a coarse approximation, stars) will have the same brightness in broad and narrow filters.}
We then produce\rr{d} a somewhat star-free F466N image, which we label 466m410, by subtracting the 410m405 image scaled by \rr{$R=(4.66/4.10)^{-2} = 0.77$ (assuming a blackbody on the Rayleigh-Jeans tail, $S_\lambda\propto\lambda^{-2}$)}.
\begin{equation}
S_{466m410} = S_{F466N} - S_{410m405} \cdot R
\end{equation}
This \rr{$S_{466m410}$}image has much greater residuals, since the wavelengths do not overlap and differences in dust extinction (and ice absorption; see below) render the subtraction somewhat poor.
Nevertheless, the stars are largely removed, and in particular, their extended PSFs are mitigated.
The subtraction process is recorded in the notebooks \texttt{BrA\_separation\_nrca.ipynb},\\
\texttt{BrA\_separation\_nrcb.ipynb},\\
\texttt{F466N\_separation\_nrca.ipynb},\\
and \texttt{F466N\_separation\_nrca.ipynb}.

This process still left significant residuals throughout both the 405m410 and 466m410 images.
To further remove stars---at this stage, purely for aesthetic purposes---we identif\rr{ied} the locations of significant residuals and masked them out, then interpolate\rr{d} across them.
We perform\rr{ed} this process iteratively, using larger masks for stars with more extended PSF features and smaller masks for more compact, fainter stars.
The details of the process were largely decided `by hand', i.e., testing a small variation in a parameter (e.g., the mask size) and revising if it did not look good.
We also created custom masks to remove residuals from extended PSFs.
The masking process is recorded in the \texttt{StarDestroyer\_nrca.ipynb} and \texttt{StarDestroyer\_nrcb.ipynb} notebooks.
After each module was fully star-subtracted, the images were merged in the \texttt{Stich\_A\_to\_B.ipynb} notebook.

Figure \ref{fig:fullfieldstarless} shows the merged full-frame image.\footnote{\url{https://www.dropbox.com/scl/fi/39cfq8yr40460qaoy1wlf/BrickJWST_merged_longwave_narrowband_lighter.png?rlkey=lusbv81fsr9rvupt99zqmvpqj&dl=0}}
\rr{This image is composed of the 466m410 image in red, 405m410 in blue, and 405m410+466m410 in green.}
\rr{We created it by building three layers, 466m410, 405m410, and (466m410 + 405m410), which we then composed into an RGB image cube.}
Figure \ref{fig:bluestarsonstars}a shows the \rr{same image}, and Figure \ref{fig:bluestarsextended}b shows the version with stars \rr{un-removed}.
\rr{Figure \ref{fig:bluestarsextended} also shows a subset of the cataloged stars in green and blue X's, as will be described in \S \ref{sec:bluestars}.}


\section{Results}
\label{sec:results}
\begin{figure*}
    \centering
    \begin{tikzpicture}
        \node[anchor=north west,inner sep=0pt] at (0,0){
            \includegraphics[width=0.49\textwidth]{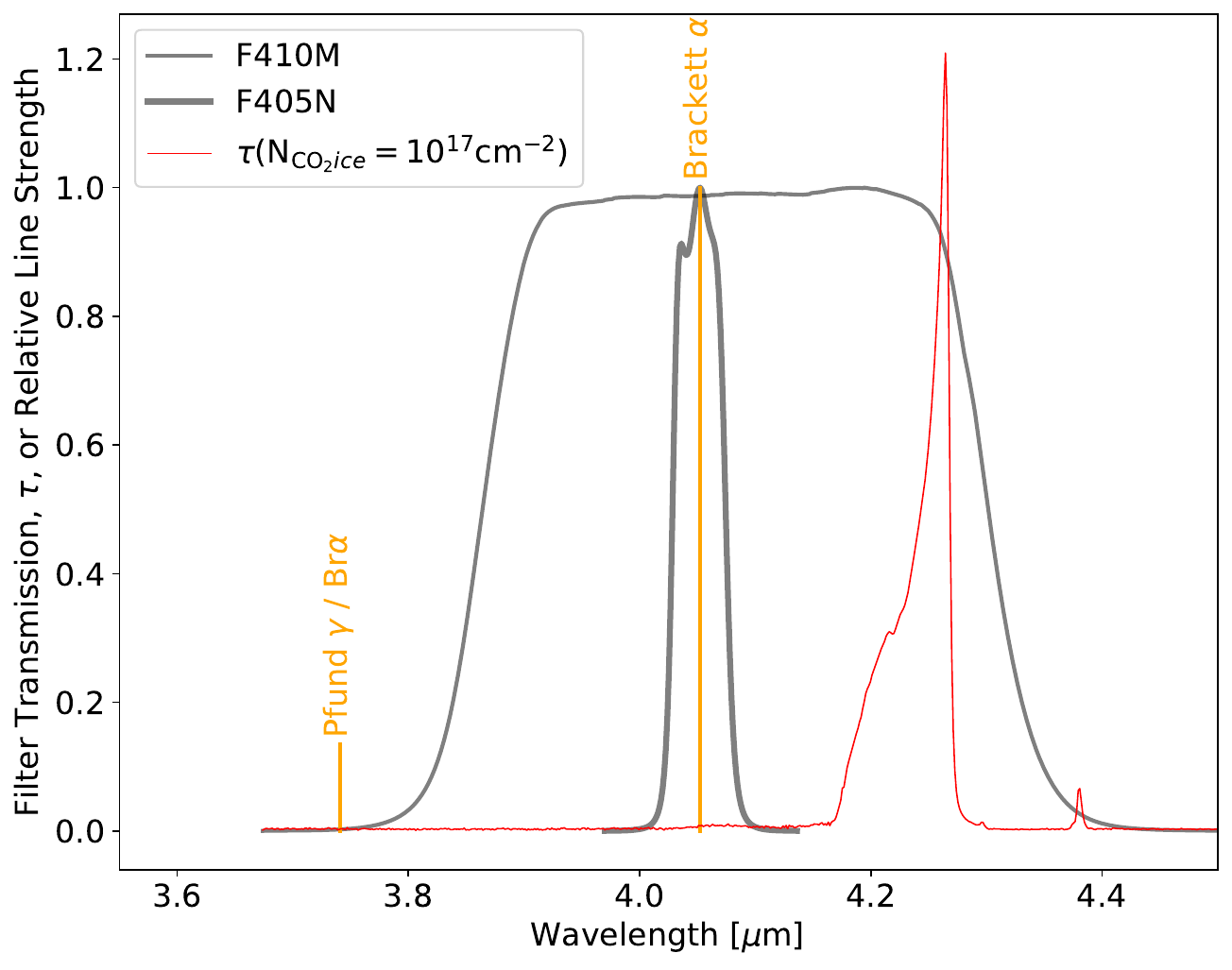}
        };
        \node[] at (8ex,2ex) {(a)};
    \end{tikzpicture}
    \begin{tikzpicture}
        \node[anchor=north west,inner sep=0pt] at (0,0){
            \includegraphics[width=0.49\textwidth]{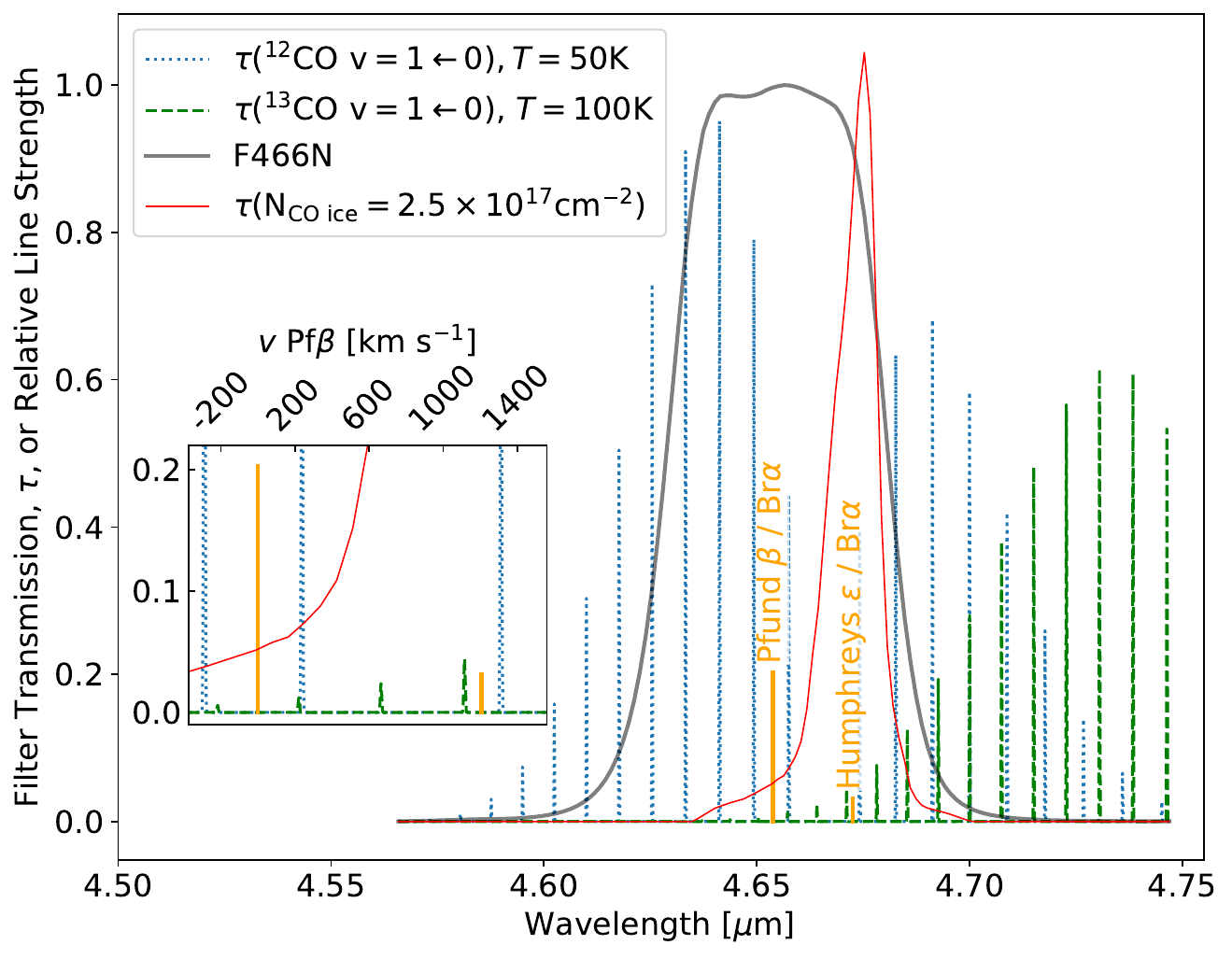}
        };
        \node[] at (8ex,2ex) {(b)};
    \end{tikzpicture}
    \caption{\rr{CO ice can absorb recombination line emission, but CO gas cannot.}
    \rr{This figure shows the} transmission curve, \rr{relative intensity} of hydrogen recombination emission lines, and \rr{optical depth of} CO gas absorption lines in \rr{selected} NIRCam long-wavelength filters.
    F410M and F405N are shown in \rr{(a)} and F466N is shown in \rr{(b)}.
    This is a schematic figure showing where lines and ice features appear and what their relative strengths are, but the amplitude scales are arbitrary\rr{: we selected column densities of CO such that the peak $\tau\approx1$, while the inferred column densities are $\sim2$ orders of magnitude greater}.
    Both the F466N and F405N filters cover hydrogen recombination lines.
    We show these as orange stick spectra (i.e., they have no width), with peak levels set as the ratio of the emissivity of the labeled line to that of Br$\alpha$ under case B conditions for $n_e=10^4$ \percc and T$=10^4$ K (see \S \ref{sec:recombination}).
    The CO gas, discussed in \S \ref{sec:cogas}, overlaps extensively with the F466N filter.
    The CO gas absorption optical depth $\tau$ is shown using a synthetic spectrum with $N(CO)=5\times10^{16}$, T=50 K, and $\sigma=5$ \kms (\S \ref{sec:cogas}).
    \rr{Both $^{12}$CO and $^{13}$CO are shown with the same column density, but we expect $N(^{13}\mathrm{CO})\approx25 N(^{12}\mathrm{CO}$) \citep[e.g.][]{Henkel1985}.}
    \rr{$^{13}$CO is shown with a greater temperature, T=100 K, because at 50 K the lines near Hu$\epsilon$ are too weak to appear in the figure.}
    \rr{As shown in the inset in figure (b),} the Pf$\beta$ line lies between the CO v=\rr{1$\leftarrow$0 J=1$\leftarrow$0 and J=2$\leftarrow$1} transitions ($\Delta v \approx 200-300$ \kms), while Hu$\epsilon$ is close to the \rr{J=0$\leftarrow$1} transition ($\Delta v\approx 100$ \kms).
    These lines would therefore would require \rr{implausibly} large broadening and/or doppler shift to overlap.
    The red \rr{thin} curves show CO and CO$_2$ ice optical depth $\tau$ using \citet{Hudgins1993} transmission curves, which overlap with the recombination lines.
    }
    \label{fig:LinesInBand}
\end{figure*}

The Brick stands out in infrared images as a dark feature against a background both of stars and of diffuse emission (e.g., Figure \ref{fig:fullfieldstarless}). 
We start by highlighting \rr{in \S \ref{sec:bluestars}} that The Brick is an extinction feature, but that it exhibits peculiar colors in the F466N filter.
We then discuss the diffuse emission from recombination lines in \S \ref{sec:recombination}.
Both from absorption of this diffuse emission and from the colors of extincted stars, we infer that gas (\S \ref{sec:cogas}) and ice (\S \ref{sec:coice}) are contributing to the line-of-sight absorption that defines The Brick.
\rr{This physical explanation is summarized in Figure \ref{fig:LinesInBand}, which shows the atomic and molecular lines and the ice bands overlaid on the observed filters.
The observational result is summarized in Figure \ref{fig:co_color}, which shows the photometric data.}


\subsection{Some stars are too blue in F466N colors}
\label{sec:bluestars}
\begin{figure*}
    \centering
    \begin{tikzpicture}
        \node[anchor=north west,inner sep=0pt] at (0,0){
    \includegraphics[width=\textwidth]{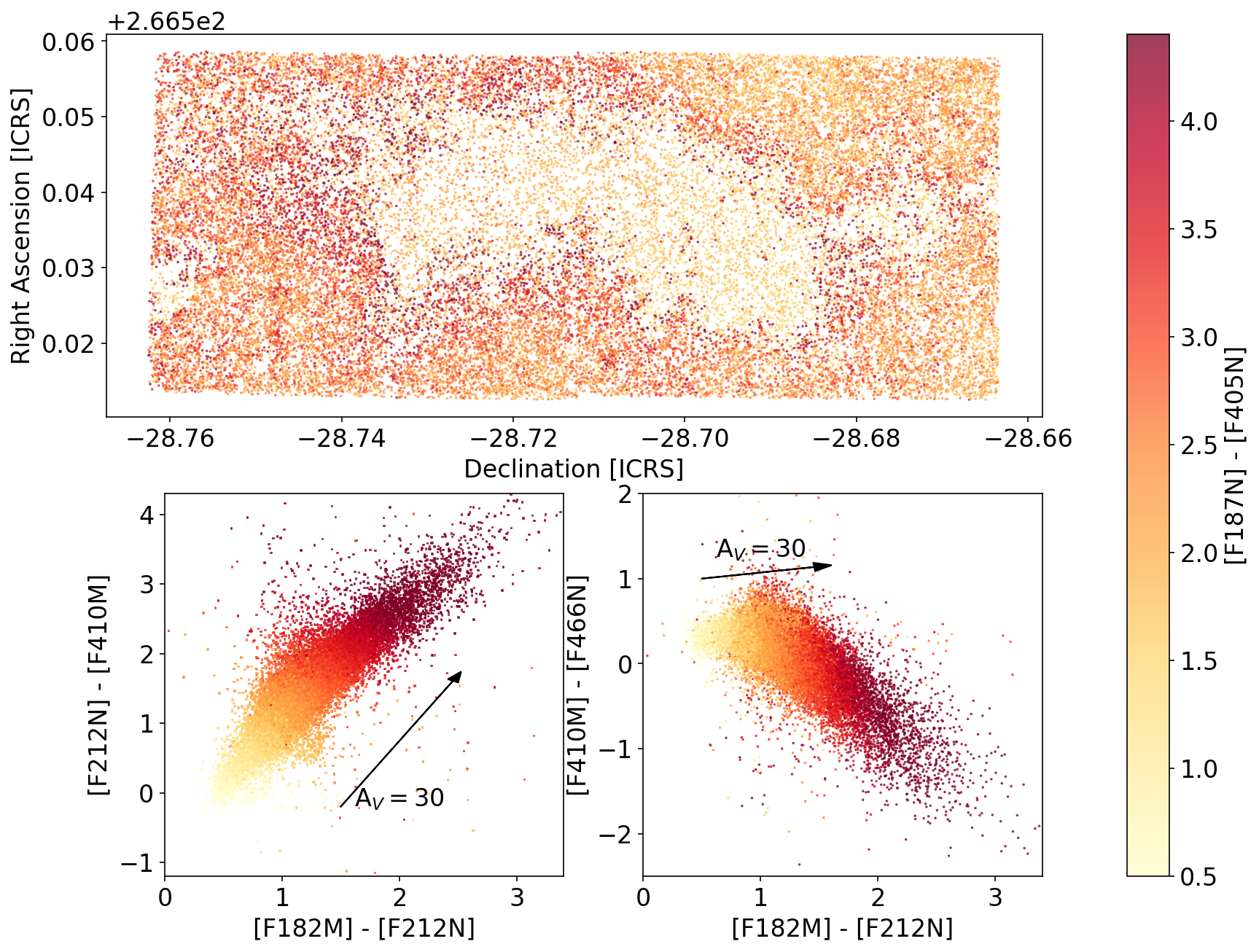}
        };
        \node[] at (13ex,-5ex) {(a)};
        \node[] at (17ex,-51ex) {(b)};
        \node[] at (65ex,-51ex) {(c)};
    \end{tikzpicture}
    \caption{
    (a, top) Location of the \nstarsccd stars with detections in all \rr{six} bands that met the quality criteria specified in \S \ref{sec:photometry} and that have [F410M]$<$\fix{15.4} \rr{mag}.
    Note that declination is on the X-axis.
    (b, lower left) Comparison of [F182M]-[F212N] to [F212N]-[F410M] color.
    (c, lower right) Comparison of [F182M]-[F212N] to [F410M]-[F466N] color.
    An extinction vector $A_V$=30 using the CT06 curve is overlaid, corresponding to about 1 magnitude of [F182M]-[F212N] color excess.
    In \rr{(b)}, the extinction vector \rr{parallels the stars in} the CCD, while in \rr{(c)}, \rr{the dust extinction vector points in a different direction: the [F410M]-[F466N] color becomes bluer, rather than redder,} with increasing extinction.
    The correlation between the colorbar and each of the colors \rr{in (b)} indicates that all of these colors align with dust extinction.
    We adopted a sequential colormap to emphasize the most extincted sources, which show up as dark red \rr{in the displayed [F187N]-[F405N] colors.
    The colormap uses [F187N]-[F405N] to demonstrate that all observed colors not involving F466N exhibit reddening consistent with dust extinction)}.
    }
    \label{fig:co_color}
\end{figure*}

The first intriguing result from these data is that, in colors including F466N, which is our longest-wavelength filter, the stars that are most extincted appear too blue.
In general, dust extinction causes reddening, i.e., the shorter-wavelength photons are more attenuated than the long-wavelength photons; in this case, we see the inverse effect happening.
This feature is evident in color-magnitude and color-color diagrams (\rr{CMDs and CCDs;} Figure \ref{fig:co_color}).
Figure \ref{fig:co_color}c shows a \rr{CCD} with an extinction vector from \citet[][hereafter CT06]{Chiar2006} overlaid\footnote{\rr{We used the \texttt{CT06\_MWGC} extinction curve from the \texttt{dust\_extinction} package because it was implemented in that package, was appropriate for the Galactic Center, and covered the range of filters used in this work.  The data used by CT06 come from \citet{Lutz1999} and \citet{Indebetouw2005}.}}, demonstrating that \rr{colors including} the F466N \rr{filter go in a direction not accounted for by normal dust-extinction-driven reddening}.
We see excess absorption in the F466N filter of roughly 0.035 magnitudes per $A_V$ with substantial scatter.

The bluest stars in [F410M]-[F466N]\footnote{\rr{We use the bracket notation, e.g., [F466N], to indicate a magnitude measurement.  [F410M]-[F466N] indicates a difference in magnitudes, i.e., a flux ratio or a color.  Negative colors are blue, positive colors are red, by convention.}} are seen at the edge of the cloud.
Figure \ref{fig:bluestarsextended} shows the location of these bluest ([F410M]-[F466N] $< -0.75$ \rr{mag}) stars.
The most extincted stars, which are the reddest in most colors but are bluest in colors involving F466N, are primarily seen along the outskirts of the cloud.
Figure \ref{fig:co_color}a shows stars color-coded by [F187N]-[F405N] color. 
\rr{We use [F187N]-[F405N] color as a consistency check: this color is very well correlated with both [F182M]-[F212N] and [F212N]-[F410M] color in panel (b), indicating that the trend seen in panel (c) is not caused by F410M.}
The interior of the cloud appears relatively blue in Fig. \ref{fig:co_color}a because only low-extinction stars are detected in the shorter-wavelength \rr{filters, and we plot only stars with detections in all six filters in this figure}.
Figure \ref{fig:co_color}b highlights that colors not involving the F466N filter are consistent with extinction.



\subsection{The F466N recombination lines \rr{are fainter than expected}}
\label{sec:recombination}
The diffuse emission circumscribing The Brick, seen in F405N and F466N in Figure \ref{fig:fullfieldstarless}, is comprised primarily of Br$\alpha$ (F405N) and Pf$\beta$ plus Hu$\epsilon$ (F466N) emission.
Figure \ref{fig:LinesInBand} shows where these lines reside with respect to the transmission profiles of the filters.
There are no other expected sources of emission in this band, as there are no known PAH features in the 4-5\um range and free-free emission is expected to be weaker by $\sim100\times$ in the narrow bands.
The ratio of hydrogen recombination lines is governed by simple rules under the assumption of Case B recombination, which is expected at moderate densities.
The expected ratio of Pf$\beta$ / Br$\alpha$ under Case B recombination at electron temperature $T_e\sim5000-10^4$ K is $R_{Pf\beta/Br\alpha} = 0.202$ \citep{Storey1995}.
The ratio of Pf$\beta$+Hu$\epsilon$, the sum of the two lines in F466N, to Br$\alpha$ is $R_{(Pf\beta+Hu\epsilon)/Br\alpha} = 0.234$.
The average foreground extinction toward the CMZ \citep{Launhardt2002,Nogueras-Lara2021} \rr{is $A_V=30$.}
Using a CT06 extinction curve, \rr{in the absence of narrow spectral features, the ratio above} rises to $R_{(Pf\beta+Hu\epsilon)/Br\alpha}(A_V=30) = 0.269$.
At \rr{greater} extinction, this ratio (\rr{which is equivalent to [F405N]-[F466N] color}) is expected to rise \rr{(become redder)}.
However, contrary to this expectation, we see the \rr{[F405N]-[F466N] color becoming more negative (bluer)} along the edge of The Brick (Figures \ref{fig:fullfieldstarless} \& \ref{fig:bluestarsextended}).
We are therefore seeing that, in \rr{regions of greater} extinction, the ratio is the inverse of what is expected from dust extinction \rr{alone}.

\subsubsection{CO absorption of the F466N recombination lines}
\label{sec:coabsorbrecomb}
There are additional absorption processes that affect only the F466N filter.
The F466N filter covers both CO gas and ice features (we will discuss these further in \S \ref{sec:cogas} \& \S \ref{sec:coice}), and therefore we expect the \rr{[F405N]-[F466N] color} to be \rr{more negative} than the theoretical Case B recombination value if CO ice is present along the line of sight.
We observe this decrease: the edges of the molecular cloud appear brown in Figure \ref{fig:fullfieldstarless}, indicating a relative deficiency in the F466N filter compared to regions f\rr{a}rther from the molecular cloud.

To assess whether the absorption is caused by CO gas or ice, we model the absorption caused by CO.
Figure \ref{fig:LinesInBand}b shows a CO line profile modeled assuming local thermodynamic equilibrium (LTE) conditions for a column density of N(CO)=$5\times10^{16}$ \persc, temperature $T=50$ K, and linewidth $\sigma=5$ \kms.
This figure shows that there is a \rr{greater than} $100$ \kms offset between the CO gas lines and the hydrogen recombination lines.
The broadest linewidths observed in the molecular gas are \rr{less than} $20$ \kms \citep{Henshaw2019}, so CO lines are unlikely to strongly absorb the recombination line emission.

By contrast, CO ice produces broadband absorption that affects both the Pf$\beta$ and Hu$\epsilon$ lines.
Figure \ref{fig:LinesInBand} shows CO and CO$_2$ absorption profiles both with assumed {N(CO) $=10^{17}$ \persc}.
The CO ice profile overlaps significantly with both recombination lines in the F466N filter.
\rr{We also model the CO$_2$ ice feature as a consistency check: if CO ice is present, CO$_2$ ice is also likely present, and therefore we verify that CO$_2$ ice would not undo the observed blue colors in [F405N]-[F466N].}
The CO$_2$ ice, shown in panel (a), significantly \rr{overlaps with the} F410M filter but has little \rr{overlap with} the Br$\alpha$ line\rr{, confirming that CO$_2$ ice can be present without driving [F405N]-[F466N] toward the red}.

Based on the observation that there is excess absorption of the diffuse F466N, \rr{and the modeling shown in Figure \ref{fig:LinesInBand},} we conclude that CO ice, and not CO gas, is absorbing Pfund $\beta$ and Humphreys $\epsilon$ emission.

\subsection{CO Gas}
\label{sec:cogas}
As shown in Figure \ref{fig:LinesInBand}, the $^{12}$CO lines in the F466N band can produce significant absorption against stellar continuum light.
In this section, we evaluate whether CO gas can produce the observed stellar colors.
We already saw in \S \ref{sec:recombination} that CO gas is unlikely to produce selective extinction of Pf$\beta$+\hue.
We find here that CO gas contributes to, but does not dominate, the total absorption in F466N.

We model this absorption as a function of temperature, column density, and linewidth assuming local thermodynamic equilibrium (LTE) conditions.
Details of this modeling are given in an associated Jupyter notebook, \texttt{COFundamentalModeling.ipynb}, that can be found in the associated GitHub repository.
The model implementation is in the \texttt{pyspeckit-models} package, which implements models compatible with \texttt{pyspeckit} \citep{Ginsburg2022}.
We used transition and level tables from the exomol database \citep{Tennyson2016} derived from \citet{Li2015} using \citet{Yurchenko2018} as an implementation reference.

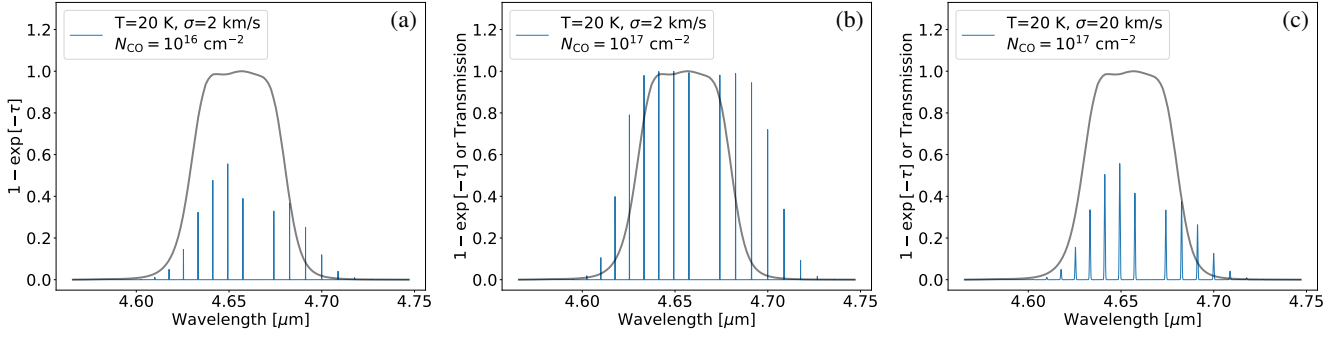
\begin{figure*}
    \centering
    
    \begin{tikzpicture}
        \node[anchor=north west,inner sep=0pt] at (0,0){
    \includegraphics[width=0.32\textwidth]{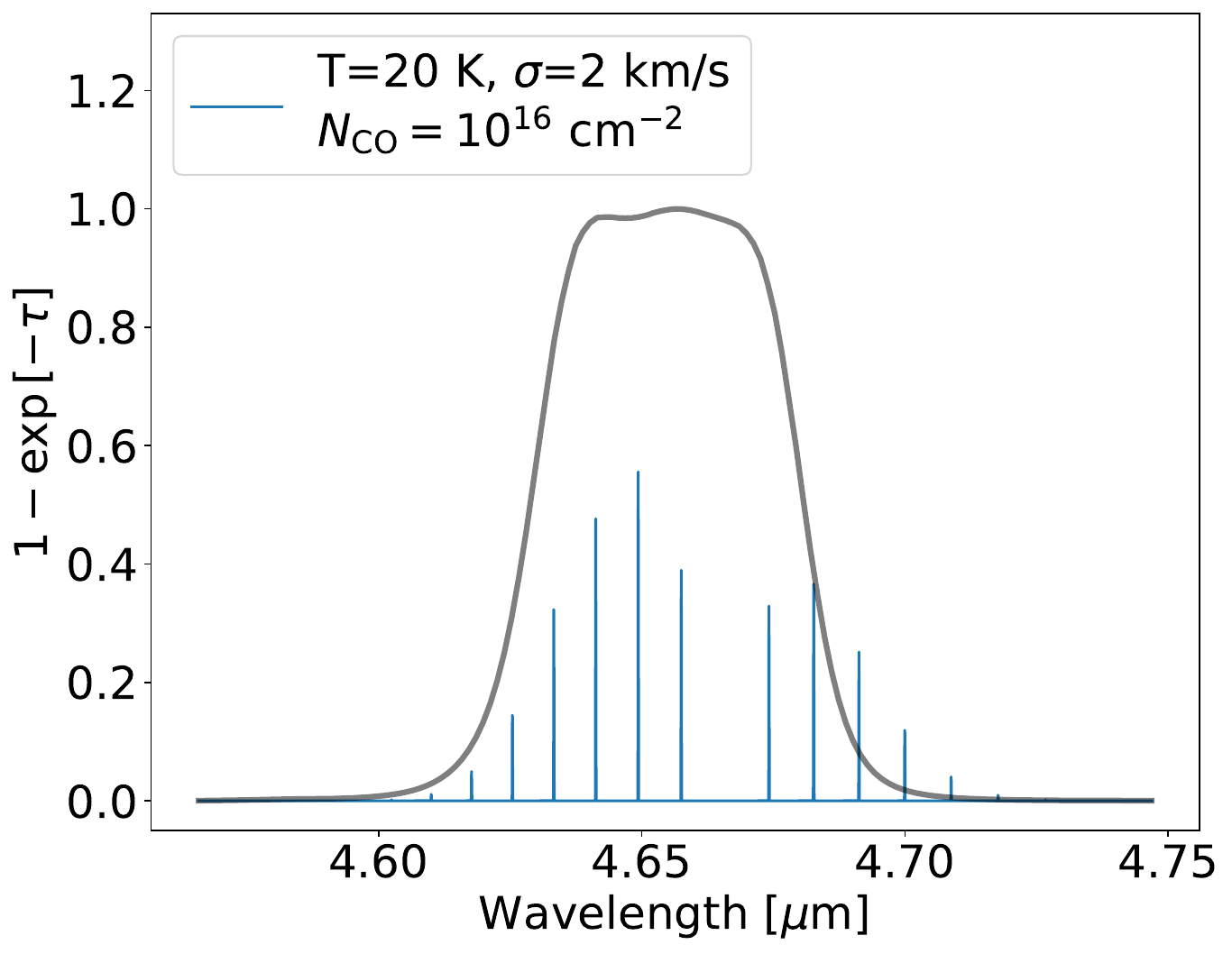}
        };
        \node[] at (37.5ex,-2.2ex) {(a)};
    \end{tikzpicture}
    \begin{tikzpicture}
        \node[anchor=north west,inner sep=0pt] at (0,0){
    \includegraphics[width=0.32\textwidth]{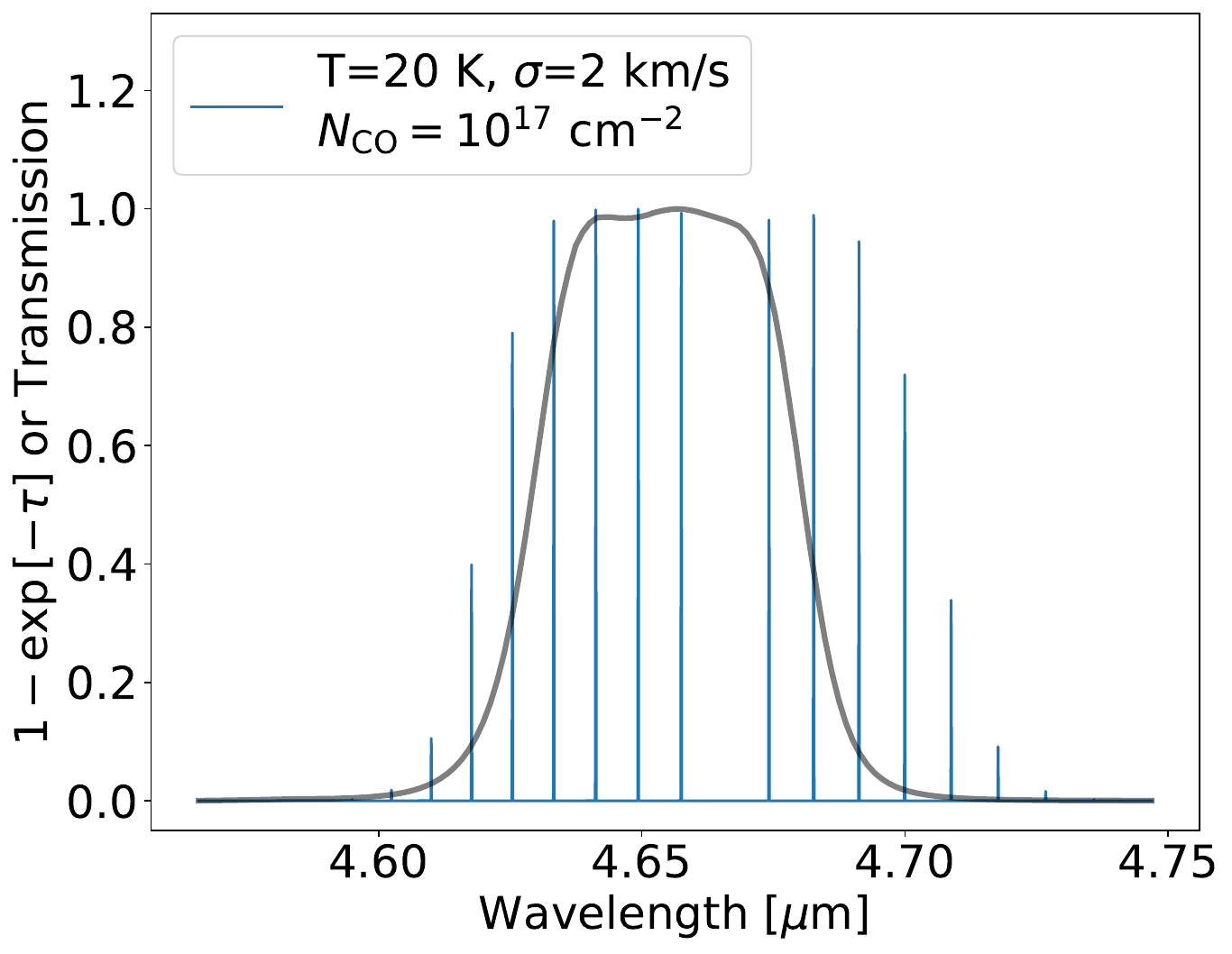}
    };
        \node[] at (37.5ex,-2.2ex) {(b)};
    \end{tikzpicture}
    \begin{tikzpicture}
        \node[anchor=north west,inner sep=0pt] at (0,0){
    \includegraphics[width=0.32\textwidth]{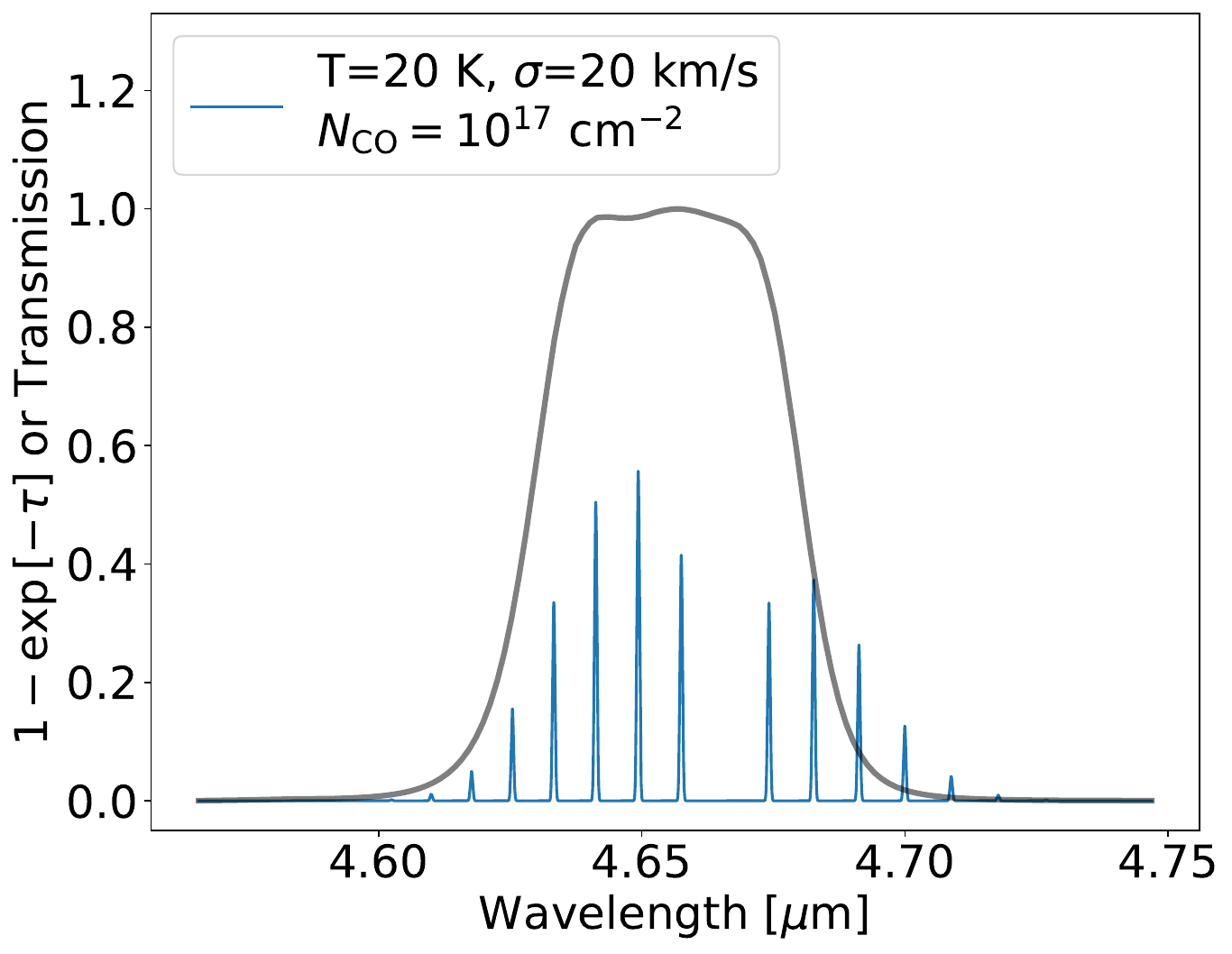}
    };
        \node[] at (37.5ex,-2.2ex) {(c)};
    \end{tikzpicture}
    \caption{Example spectra showing the optical depth of \rr{gas-phase} $^{12}$CO v=\rr{1$\leftarrow$0 absorption} superposed on the transmission spectrum of the F466N filter for a variety of physical conditions.
    The \rr{thick} black curve in each case shows the transmission function of the F466N filter.
    The \rr{thin} blue curves show the absorbed light fraction $1-e^{-\tau}$.  
    \rr{Panel (a)} shows the effect of this absorption for a low column density case (N(CO) = $1\ee{16}$ \persc or N(H$_2$) $=1\ee{20}$ \persc for a typical CO abundance of $X_{CO} = 10^{-4}$).
    \rr{Panel (b)} shows N(CO)=$10^{17}$ \persc (N(H$_2$) $=10^{21}$ \persc).
    \rr{Panel (c)} shows transmission for CO column density $10^{17}$ \persc (N(H$_2$) = $1\ee{21}$ \persc) but with a broader line, illustrating that more of the band is absorbed but the peak optical depth is lower.
    \rr{We model low column densities to illustrate the behavior that affects the majority of stars in our sample, which are at relatively low column density, and to highlight the effects of gas temperature on the lines that are excited.}
    \rr{The modeled column densities here fall significantly short of the peak column in The Brick, N(H$_2$)$\approx5\ee{23}$, but panel (b) captures the behavior at greater column densities: the line optical depths are large, so the lines are saturated.
    At greater column densities, there is very modest optical depth broadening, but qualitatively even extremely high column density resembles that model spectrum.}
    }
    \label{fig:COexamplespectra}
\end{figure*}

Figure \ref{fig:COexamplespectra} shows example optical depth spectra overlaid on the transmission profile of the F466N filter.
\rr{The v=1$\leftarrow$0 and J=4$\leftarrow$3, 3$\leftarrow$2, 2$\leftarrow$1, 1$\leftarrow$0 R-branch transitions and the v=1$\leftarrow$0 J=0$\leftarrow$1 
 P-branch} transition all lay within the range \rr{where} F466N has $>50$\% of peak transmission \rr{(we use $\leftarrow$ in the transition names to indicate that these are absorption lines)}.
The maximum absorption in this filter occurs for temperatures between 10-20 K, while we expect \rr{gas} temperatures near 50-100 K \citep{Ginsburg2016,Immer2012,Krieger2017}.
At \rr{greater} temperatures, a large fraction of CO molecules are in states that only produce transitions outside of the F466N band, reducing the absorption \rr{for a fixed assumed column density}.
It is \rr{possible} that \rr{some} of the CO gas is at moderate densities ($n(H_2)\lesssim10^4$ \percc) and therefore is sub-thermally excited, which would concentrate the CO molecules into the lower-J levels, \rr{thereby reducing the effect of high gas temperature}.
Nevertheless, the LTE models shown in Figure \ref{fig:COexamplespectra} capture the range of expected behavior.

We model the CO gas absorption for the expected range of line width and column density in the Galactic Center.
For narrow line width\rr{s}, such as \rr{those} caused by thermal broadening at $T<100$ K, the absorption is negligible.
In the Galactic Center, there is significant doppler broadening that is generally attributed to turbulence.
The total linewidth in the cloud may range from $\sigma\sim5-20$ km/s \citep{Henshaw2019}.
CO column densities will span the full range from effectively zero (since CO is destroyed by UV at $A_V \lesssim 2$) to $\sim$ a few$\times10^{19}$ cm$^{-2}$ \citep{Rathborne2015} assuming $X_{CO}=10^{-4}$.
In our observations, we detect stars \rr{at wavelengths} short of 2\um only at intermediate column densities, most likely below $A_V<80$ mag ($N(H_2) < 10^{23}$ \persc), since dust extinction hides the stellar continuum at \rr{greater} column density \rr{at the current level of sensitivity}.

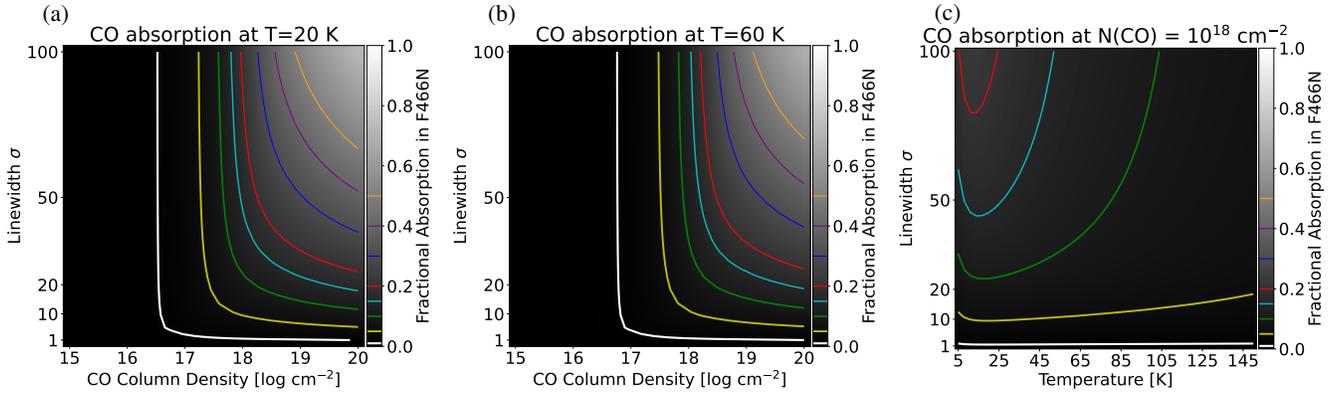
\begin{figure*}
    \centering
    \begin{tikzpicture}
        \node[anchor=north west,inner sep=0pt] at (0,0){
    \includegraphics[width=0.32\textwidth]{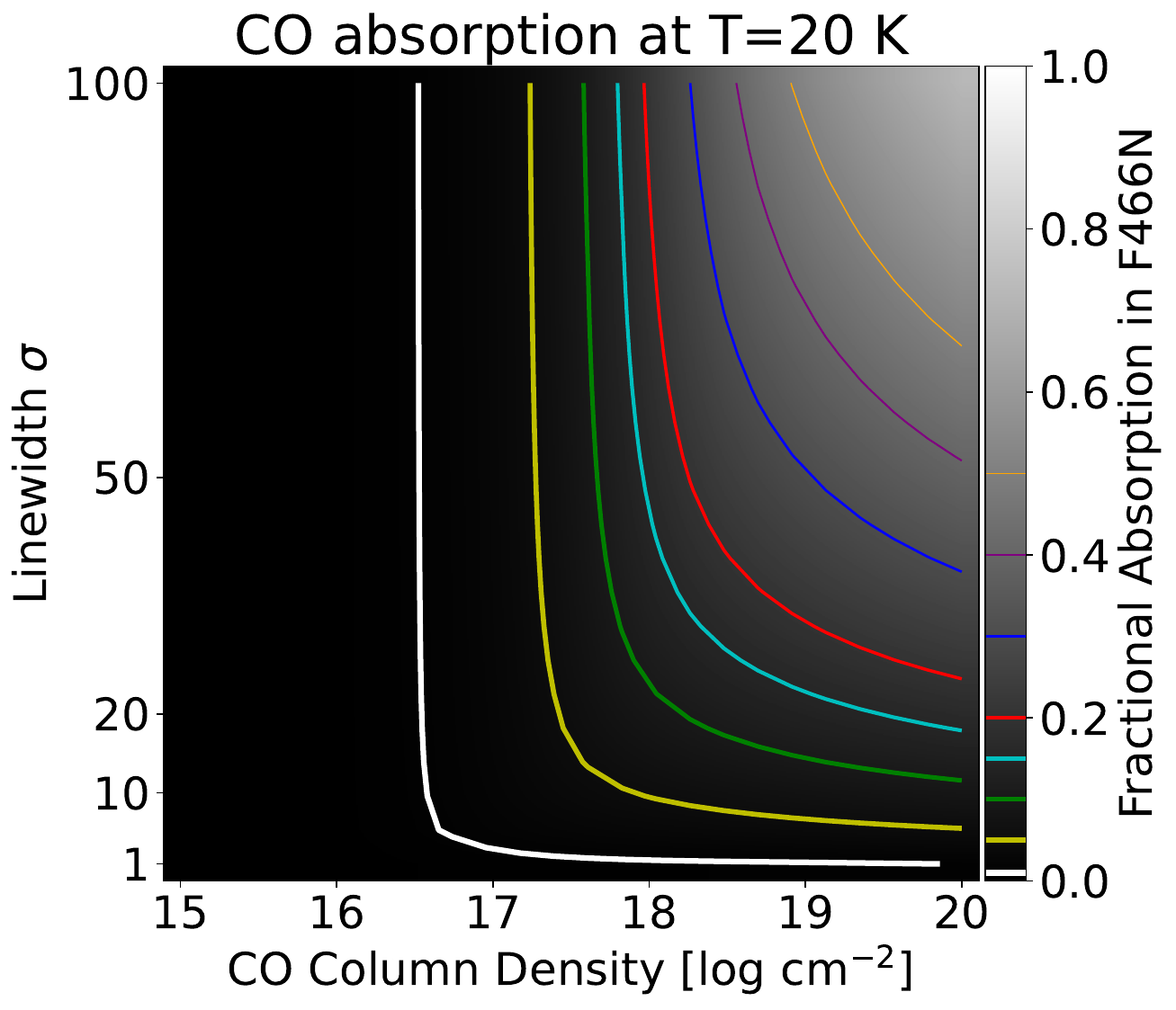}
    };
        \node[] at (5ex,0.5ex) {(a)};
    \end{tikzpicture}
    \begin{tikzpicture}
        \node[anchor=north west,inner sep=0pt] at (0,0){
    \includegraphics[width=0.32\textwidth]{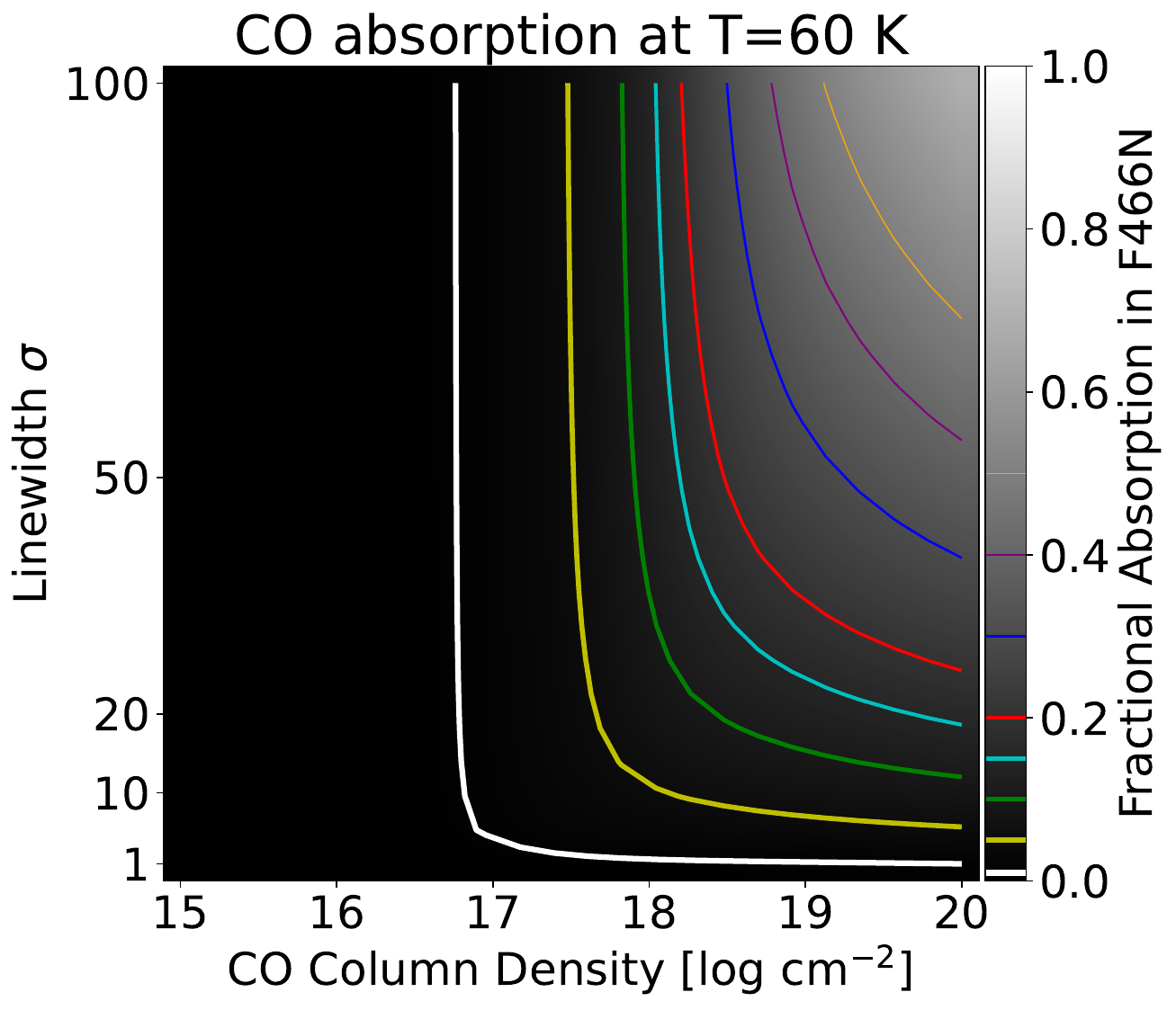}
    };
        \node[] at (5ex,0.5ex) {(b)};
    \end{tikzpicture}
    \begin{tikzpicture}
        \node[anchor=north west,inner sep=0pt] at (0,0){
    \includegraphics[width=0.32\textwidth]{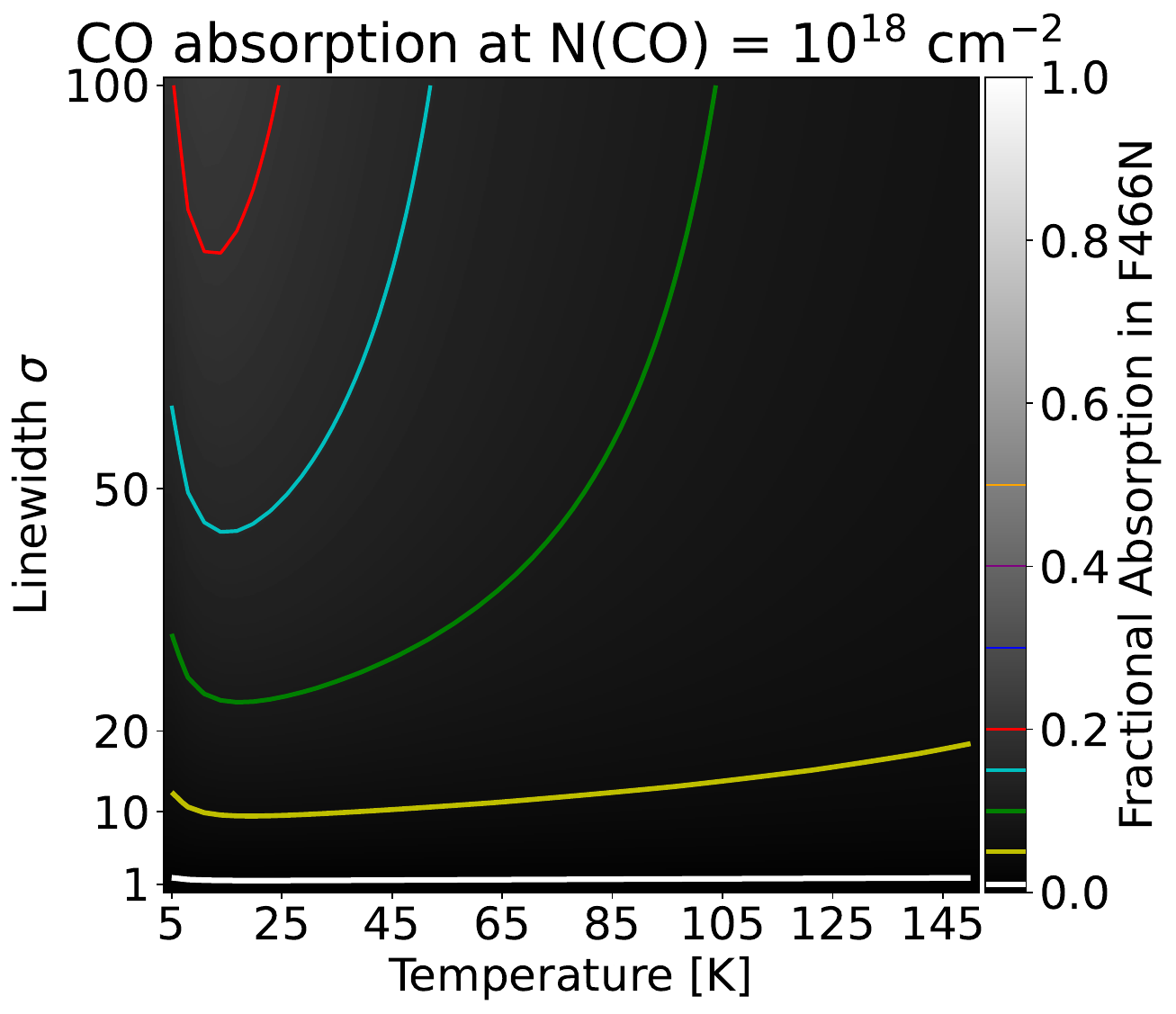}
    };
        \node[] at (5ex,0.5ex) {(c)};
    \end{tikzpicture}
    \caption{The fractional absorption by gas-phase CO v=\rr{1$\leftarrow$0 absorption} within the F466N band for 20K (left), 60 K (middle), and at a fixed column density $N_{CO}=10^{18}$ \persc (right).
    The grayscale shows the fractional absorption as labeled in the colorbar; an absorption of 1.0 implies that no photons are received in the band.
    Colored lines are placed at levels=[0.01, 0.05, 0.1, 0.15, 0.2, 0.3, 0.4, 0.5] with colors=[white, yellow, green, cyan, red, blue, purple, orange], respectively, to help guide the eye.
    \rr{These curves occur in order from bottom to top and the line widths progress from thick to thin.}
    These levels correspond to magnitude differences $\Delta m=$ [0.011, 0.056, 0.114, 0.176, 0.242, 0.387, 0.555, 0.753].
    }
    \label{fig:COmodelgrids}
\end{figure*}

Figure \ref{fig:COmodelgrids} summarizes the modeling results.
Given the plausible range of column density and line width,
the total CO gas absorption in F466N may range from $\sim1\%$ to at most $\lesssim20\%$.

At the column densities where the absorption is easily detectable (fractional absorption $\gtrsim0.1$ results in a change in magnitude $\Delta m \gtrsim 0.1$; green line in Figure \ref{fig:COmodelgrids}), change in linewidth dominates over the change in column density or temperature.
The foreground Galactic disk clouds, which have narrow lines, produce relatively little absorption; we therefore argue that intervening material between us and the Galactic Center is not primarily responsible for \rr{the blue [F410M]-[F466N] color of} the stars.
These models also imply that, even at very extreme column densities ($N_{CO}>10^{19}$ \persc), ``normal'' galactic disk clouds with $\sigma<10$ \kms will produce minimal CO absorption in the F466N band, while typical galaxy center and galactic bar clouds with $\sigma\sim10-50$ \kms will produce readily detectable absorption.
However, even for very broad lines ($\sigma=50\kms$) at high column ($N_{CO}\sim10^{19.5}$ \persc), CO gas produces $\lesssim1$ magnitude ($<50\%$) of absorption \rr{in the F466N band}.


\subsection{CO ice absorption}
\label{sec:coice}
While we show above that CO gas can produce a substantial amount of absorption, the observed absorption depths reach levels difficult to explain with gas alone. 
Figure \ref{fig:COmodelgrids} shows that CO column densities $N(CO)>10^{19.5}$ \persc, implying $N(H_2)>3\times10^{23}$ \persc, are required to explain the $\gtrsim2$ magnitudes of F466N absorption shown in Figure \ref{fig:co_color}.
Such high column densities are rare in The Brick \citep{Rathborne2014a}, occurring primarily in the inner regions (see their Figure 2) and in dense cores \citep{Walker2021}, while the high-extinction and high-CO-absorption stars we detect are primarily in the outskirts (Figure \ref{fig:bluestarsextended}).
Stars behind these high column densities would be too extincted to detect in the shorter wavelength band; the highest extinction we report in \rr{\S \ref{sec:coicevsext}} is $A_V\sim80$ \rr{mag}, or $N(H_2)\sim10^{23}$ \rr{\persc}.
Additionally, in Section \ref{sec:recombination}, we showed that CO gas is unlikely to absorb recombination lines.
We therefore examine the possibility that ice absorption is responsible for the observed F466N \rr{deficits}.

There is evidence that The Brick contains some ice, but \rr{also} that CO is not entirely frozen out.
Pure CO ice forms at low temperatures, $T<20$ K \citep{Hudgins1993}.
The average dust temperature in The Brick is close to 20 K \citep{Tang2021}, so it is probable that some of the volume of The Brick is cold enough to freeze CO.
The Brick exhibits signs of substantial freezeout in its center based on gas observations \citep{Rathborne2014b}, but still has substantial gas-phase CO detected \citep{Ginsburg2016,Rigby2016,Eden2020}.
It is likely that much of the observed gas-phase CO is on the cloud surface, while further into the interior, CO is more completely frozen out.

\subsubsection{CO ice modeling}

To model CO ice absorption, we convolve\rr{d} \rr{a given} filter transmission curve with a synthetic \rr{stellar} model spectrum.
We \rr{began} with a 4000 K PHOENIX stellar atmosphere \citep{Husser2013} as the base model, then examine\rr{d} the fractional flux lost in the F466N band as a function of CO column density.
We retrieve\rr{d} optical constants for pure CO ice and CO mixed with OCS and CH$_4$ in a 20:1 ratio from  \citet{Hudgins1993} via the JPL Optical Constants Database\footnote{\url{https://ocdb.smce.nasa.gov/page/toc}}.
\rr{We do not know which ice mixture is most appropriate for our data, so we chose to show all available laboratory mixtures in which CO was the primary constituent.}
We also considered the possibility that CO is embedded in other ices \citep[e.g., \water and \methanol][]{Pontoppidan2003,Boogert2008}, but found little practical difference from pure CO ice when using the \citet{Hudgins1993} and \citet{Rocha2016} optical constants.

\begin{figure}
    \centering

    \begin{tikzpicture}
        \node[anchor=north west,inner sep=0pt] at (0,0){
    \includegraphics[width=0.465\textwidth]{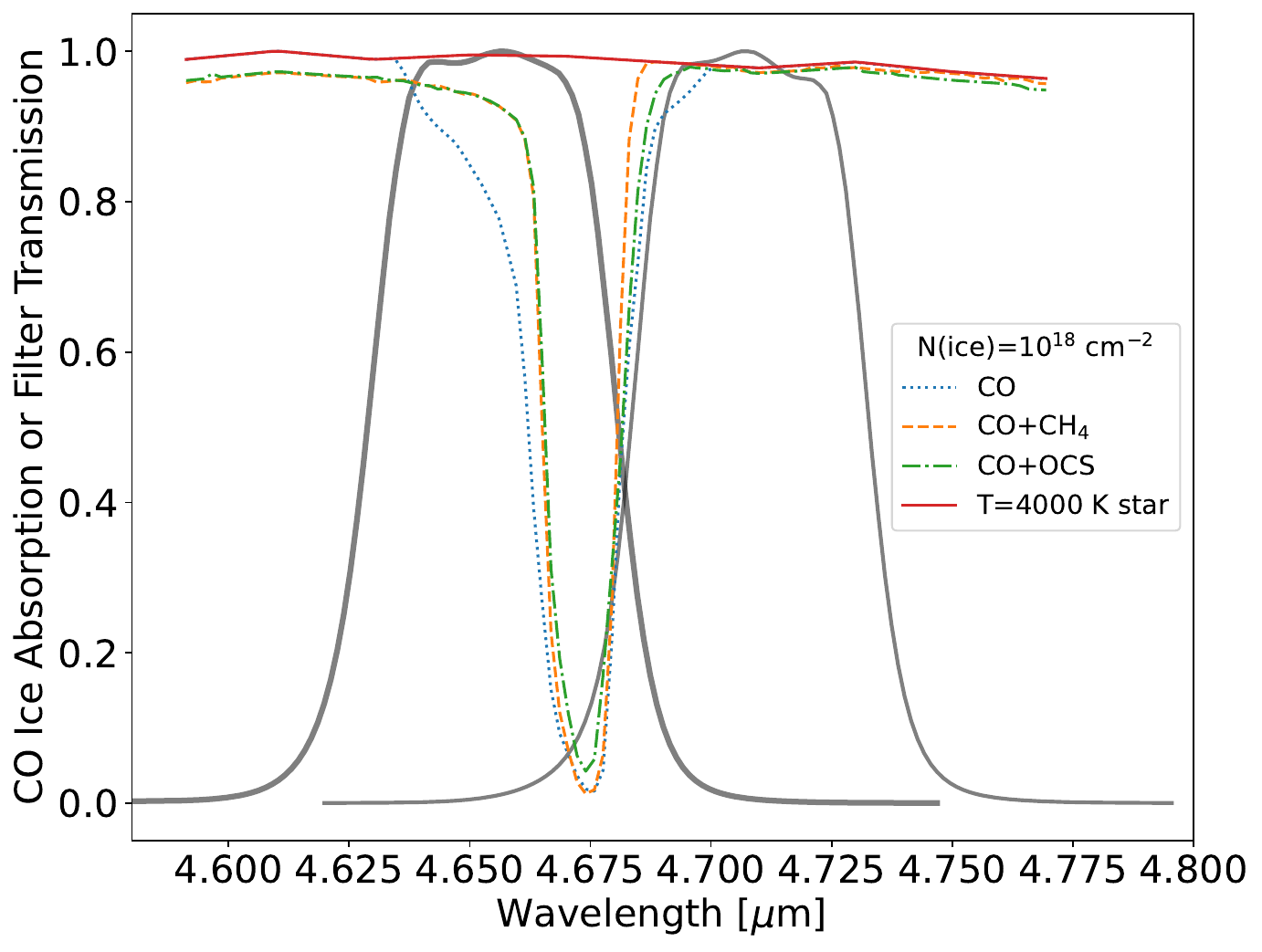}  
    };
        \node[] at (8ex,1ex) {(a)};
    \end{tikzpicture}
    \begin{tikzpicture}
        \node[anchor=north west,inner sep=0pt] at (0,0){
    \includegraphics[width=0.515\textwidth]{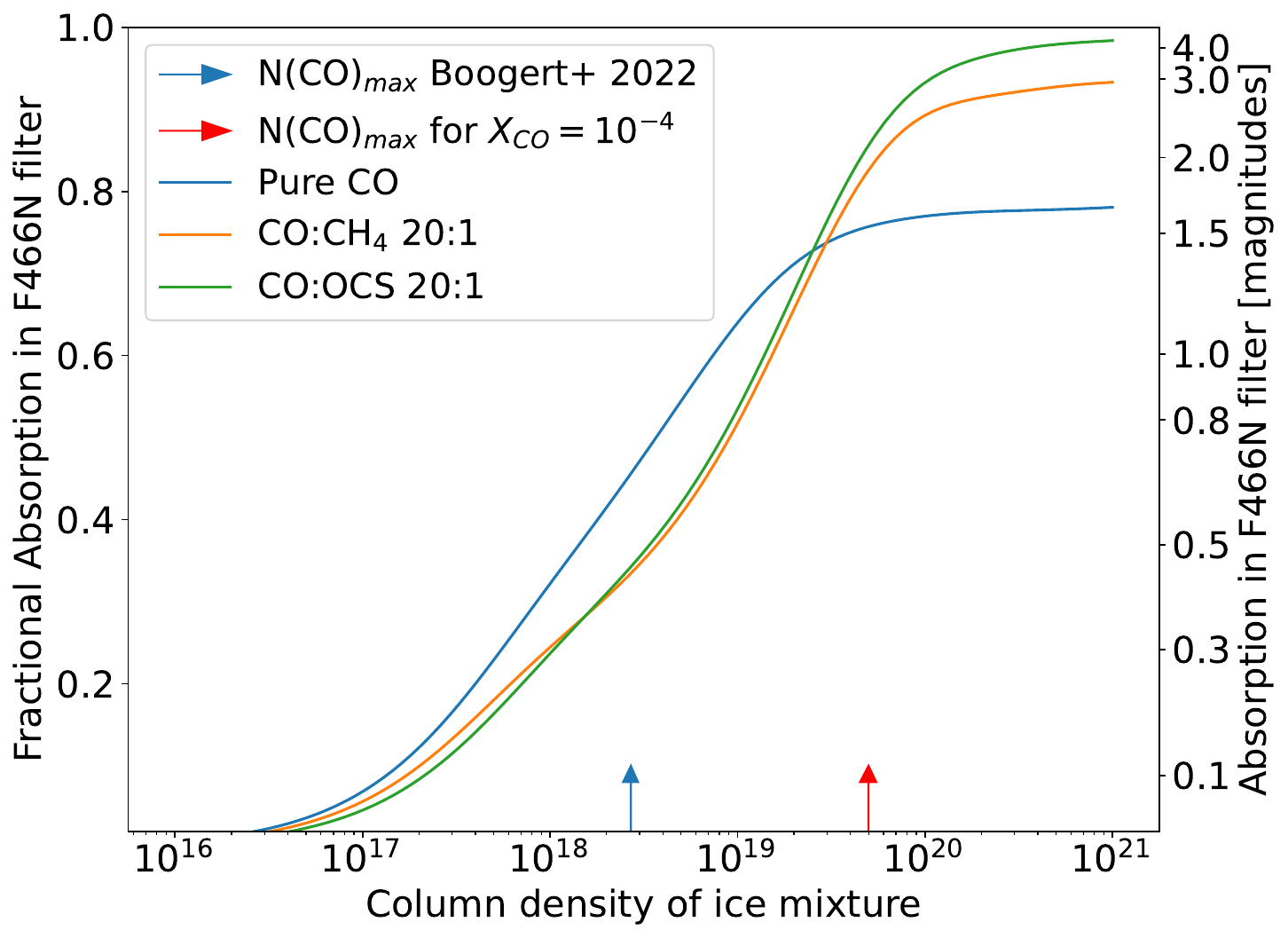}
    };
        \node[] at (8ex,1ex) {(b)};
    \end{tikzpicture}
    \caption{
    (\rr{a}) Overlap of the transmission profiles of the F466N and F470N filters with CO \rr{ice absorption models superimposed on a PHOENIX 4000 K stellar photosphere model.}
    The absorption profiles are produced using \citet{Hudgins1993} opacity  measurements.
    (\rr{b}) The expected absorption as a function of column density in the F466N band for three different CO ices.  
    The absorption is given in fractional value on the left and magnitudes on the right.
    The arrow\rr{s} show the highest observed column density of CO ice from the \citet{Boogert2022} sample of high-mass young stellar objects \rr{and the maximum possible CO column density assuming a CO abundance with respect to H$_2$ of 10$^{-4}$ given the observed column density in The Brick (but see \S \ref{sec:wrapupdiscussion} for a discussion of this assumption).}
    }
    \label{fig:icemodel}
\end{figure}

Figure \ref{fig:icemodel} shows the effects of \rr{CO} ice absorption\footnote{Figure \ref{fig:icemodel}a includes the F470N filter, which we have not used in this work, to caution other JWST users that there may be significant, albeit weaker, CO absorption in this filter.}: at $\mathrm{N(CO)}\approx10^{19}$ \persc, we expect $\approx0.5-1$ mag of absorption from the ice band.
The \rr{greatest} observed column density within The Brick, based on ALMA dust emission observations with $\sim3\arcsec$ resolution, is $\mathrm{N}(\mathrm{H}_2)\sim5\times10^{23}$ \persc \citep{Rathborne2014a}, which implies an upper limit on the ice column density $\mathrm{N(CO)}<5\times10^{19}$ \persc if we assume the CO/H$_2$ ratio is $10^{-4}$ and \rr{the} gas-to-dust mass ratio \rr{is} 100.
If \emph{all} of the CO is frozen out into pure CO ice, in the highest column-density line-of-sight, the absorption \rr{in F466N} could just about reach 1.2 mag ($\sim65$\%).

\begin{figure}
    \centering
    \begin{tikzpicture}
        \node[anchor=north west,inner sep=0pt] at (0,0){
    \includegraphics[width=0.465\textwidth]{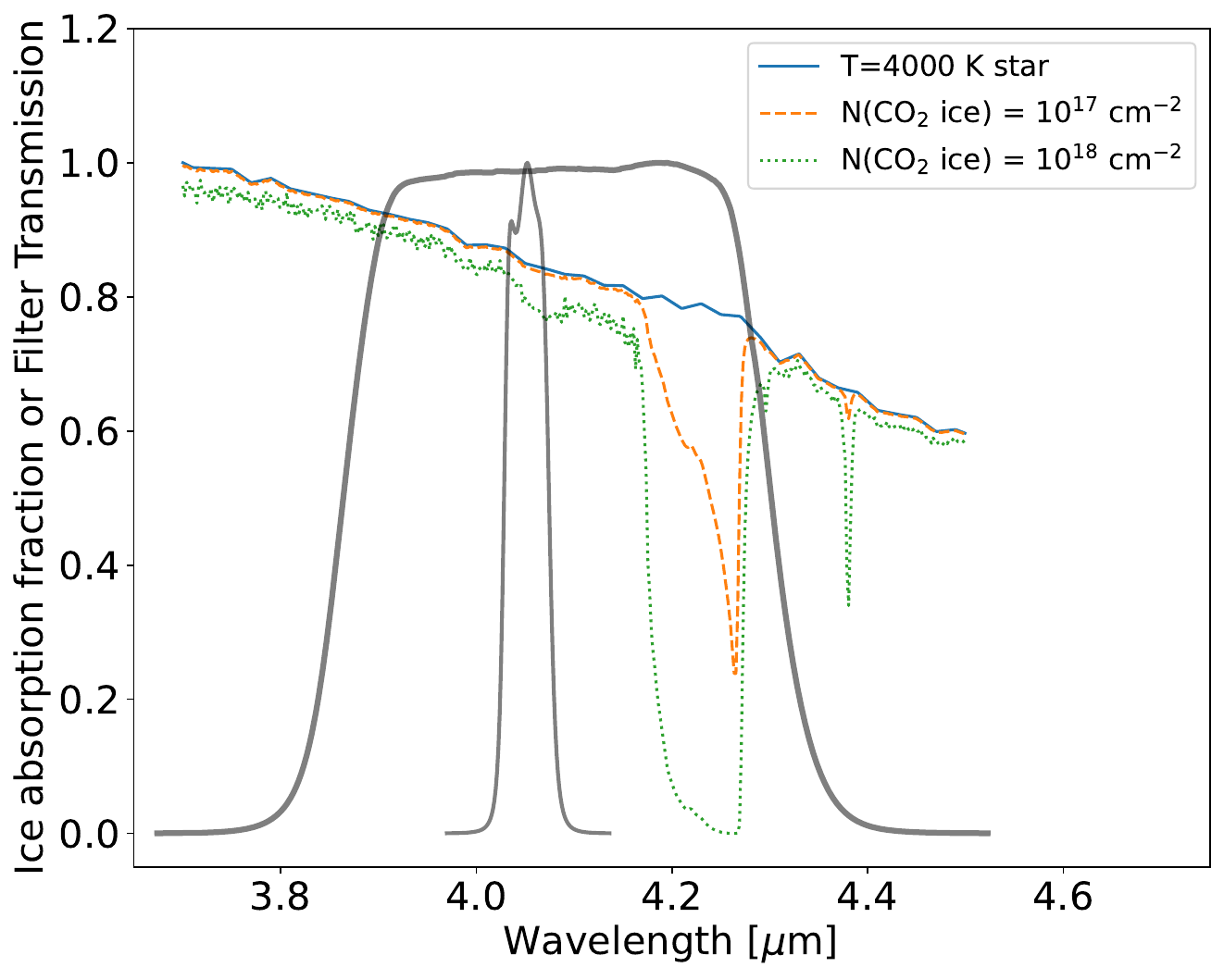}
    };
        \node[] at (8ex,1ex) {(a)};
    \end{tikzpicture}
    \begin{tikzpicture}
        \node[anchor=north west,inner sep=0pt] at (0,0){
    \includegraphics[width=0.515\textwidth]{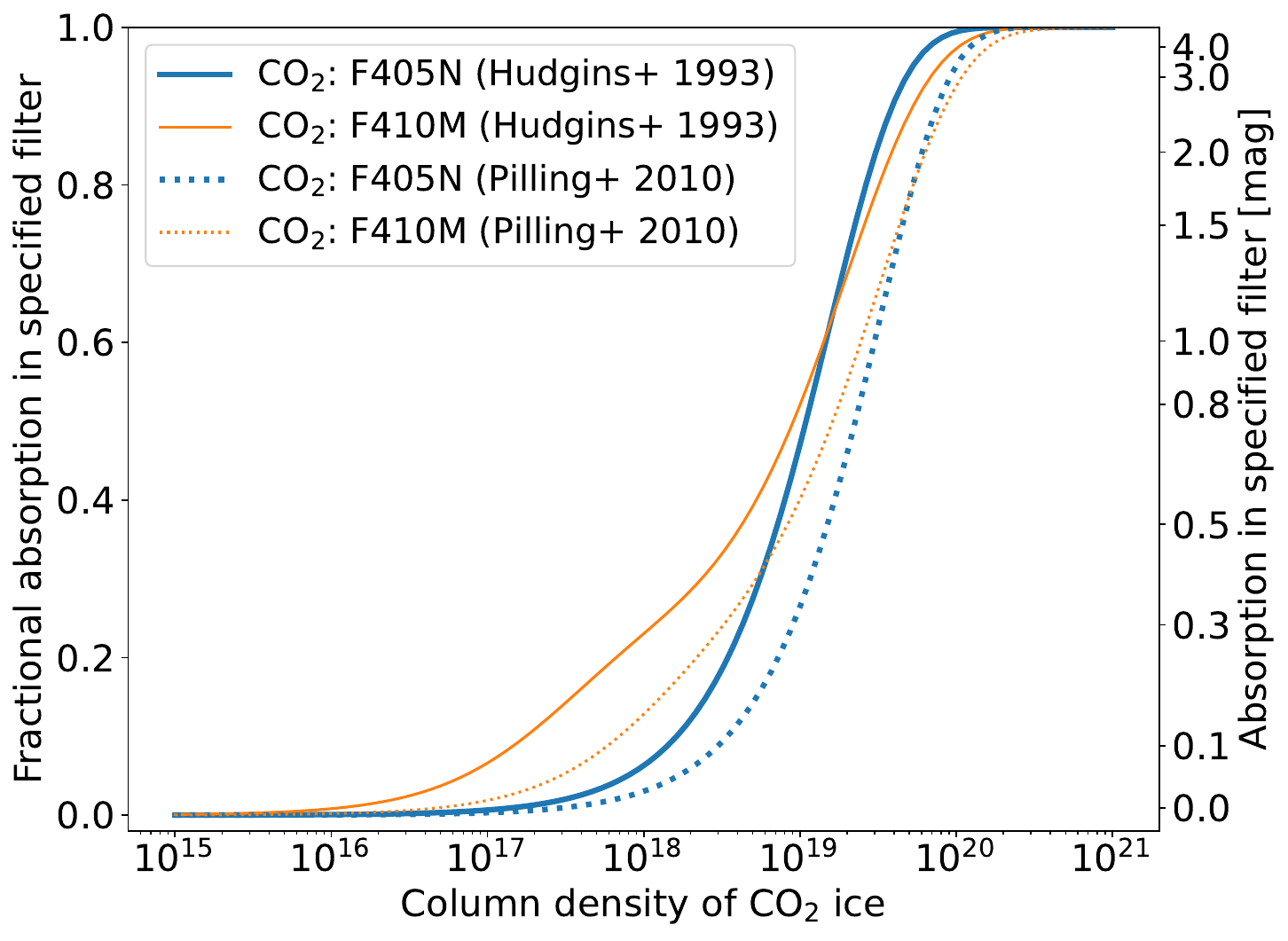}
    };
        \node[] at (8ex,1ex) {(b)};
    \end{tikzpicture}
    \caption{
    (\rr{a}) Overlap of the transmission profiles of the F405N and F410M filters with CO$_2$ ice absorption models superimposed on a PHOENIX 4000 K stellar photosphere model.
    The absorption constants are for pure CO$_2$ ice from \citet{Hudgins1993} via the Optical Constants Database (OCDB).
    (\rr{b}) The expected absorption as a function of column density in the F405N and F410M bands.
    The absorption is given in fractional value on the left and magnitudes on the right.
    }
    \label{fig:co2icemodel}
\end{figure}

Figure \ref{fig:co2icemodel} shows similar plots for the F405N and F410M filters \rr{for} CO$_2$ ice to demonstrate that CO$_2$ ice can have some effect, but \rr{less than CO}, on our observed colors.



\subsubsection{CO ice as a function of extinction}
\label{sec:coicevsext}
We compare\rr{d} the inferred extinction to the CO ice column to obtain a coarse estimate of how CO \rr{ice} varies with $A_V$.
We use [F182M]-[F212N] color to estimate $A_V$ using the CT06 extinction curve, which we justify by comparing to ground-based GALACTICNUCLEUS colors in Appendix \ref{appendix:galnuc}.

\begin{figure}
    \centering
    \includegraphics[width=0.5\textwidth]{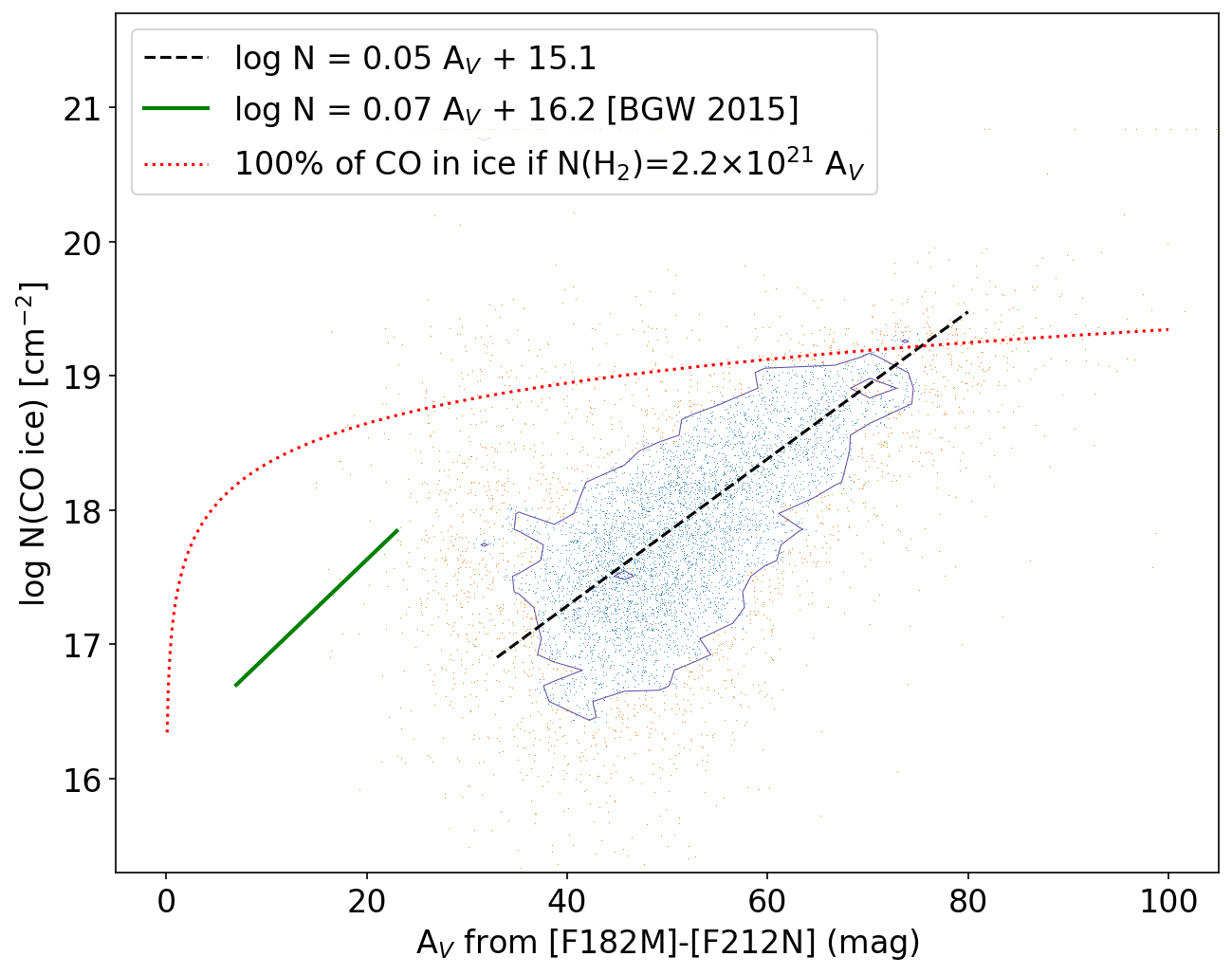} 
    \caption{N(CO ice) vs $A_V$.
    The black dashed line is shown as a representative typical value; it is not a fit to the data.
    The green \rr{solid} line is from \citet[][Figure 7]{Boogert2015}.
    The red \rr{dotted} line shows the CO column density obtained by converting N(CO) = N(H$_2$)/10$^4$.
     $A_V$ is computed from [F182M]-[F212N] color \rr{using the CT06 extinction law}.
     \rr{We caution that the X-axis is subject to systematic uncertainty from the choice of extinction curve; using the \citet{Fritz2011} curve instead of CT06 would compress the X-axis by a factor of 1.8.}
     \rr{Additionally, N(CO ice) is calculated assuming 100\% of the F466N deficit is produced by ice, not gas, while gas certainly contributes to the absorption, especially at lower column density (see \S \ref{sec:cogas}).}
     \rr{Statistical errors are limited to $\sigma(\log N(CO))\lesssim0.17$ dex and $\sigma(A_V)\leq4.4$ mag, which come from the selections imposed on the catalog (we include only stellar measurements with $\sigma<0.1$ mag) and a simple linear propagation of error for N(CO).}
    }
    \label{fig:ncovsav}
\end{figure}

To measure the CO absorption, we use\rr{d} the [F410M]-[F466N] color.
Figure \ref{fig:co_color}c shows that the [F410M]-[F466N] color \rr{becomes} bluer at \rr{greater} extinction.
We calculate\rr{d} the absorption from pure CO ice in the F466N band to obtain a mapping from N(CO) to [F410M]-[F466N] color.  
We deredden\rr{ed} our measured [F410M]-[F466N] color using the $A_V$ computed above with the CT06 extinction curve.
\rr{Note that the assumed extinction curve introduces a strong systematic uncertainty: using a \citet{Fritz2011} extinction curve would reduce the estimated $A_V$ by a factor of 1.8.}
Figure \ref{fig:ncovsav} shows the resulting N(CO) as a function of $A_V$.
\rr{Note that this curve assumes that 100\% of the deficit in the [F466N] band is produced by ice, which is an upper limit and likely overestimate because CO gas (\S \ref{sec:cogas}) also contributes, especially at low column density.}


The red curve in Figure \ref{fig:ncovsav} shows the maximum possible CO at each $A_V$ adopting standard values of  N(H$_2)=2.21\times10^{21} A_V$ \citep{Guver2009} and CO abundance relative to hydrogen $X_{CO}=10^{-4}$.
Some of the data \rr{points reside} above this \rr{curve}, suggesting that one or both of these assumptions may be incorrect, which we will evaluate further in \S \ref{sec:wrapupdiscussion}.

While there is overall correlation between N(CO) and $A_V$, the dispersion at any given $A_V$ is large, \rr{greater than one} order of magnitude.
Such a large scatter indicates a wide range of conditions, with many lines-of-sight containing little CO at the lowest observed $A_V$.


\section{Discussion}
\label{sec:icegasdiscussion}

\subsection{CO produces F466N absorption}
\label{sec:wrapupdiscussion}
The CO ice models can qualitatively explain most of the observed F466N \rr{deficits}.
However, neither the gas nor ice models appear to quantitatively explain the most deeply absorbed sources.
In both the gas and ice absorption cases, it is possible to achieve substantial absorption in the F466N band, enough to be easily detected, but less than the 1-2 mag seen in Figure \ref{fig:co_color} if typical CO abundance and gas-to-dust ratios are used.

One possible explanation \rr{lies} in our assumptions about how to convert column densities between dust, molecular hydrogen, and CO: if the CO/H$_2$ ratio is \rr{greater} than the assumed 10$^{-4}$, or the gas-to-dust ratio is \rr{less} than 100 \citep[e.g.][]{Giannetti2017}, the total CO column could be \rr{greater}.
Increasing the CO abundance or decreasing the gas-to-dust ratio\rr{, both of which are plausible because gas in the CMZ is higher-metallicity than the solar neighborhood,} would both have the effect of shifting the red dashed line upward in Figure \ref{fig:ncovsav}.
These changes would therefore increase the maximum allowed CO abundance and explain the large observed F466N absorption.


\subsection{Ice freezeout and gas thermodynamics}
\label{sec:thermodynamics}

The presence of significant quantities of CO ice in The Brick highlights the fact that the dust is significantly colder than the gas.
Gas temperatures in the CMZ generally \citep{Ginsburg2016,Krieger2017}, and The Brick specifically \citep{Johnston2014}, are observed to be high, $T\gtrsim50$ K, and in many locations $T>100$ K.
Freezing of pure CO into ice is expected to occur at dust temperatures $T\lesssim20$ K, though CO can be integrated into \water and \methanol ices that freeze out at \rr{greater} temperatures  \citep[$T\gtrsim80$ K;][]{Boogert2015,Garrod2006}.
The dust temperatures observed in The Brick have been in the range $T\sim20-30$ K, albeit at lower resolution \citep{Marsh2016,Tang2021}, which is somewhat too warm for pure CO freezeout but cold enough to freeze other molecules.
However, dust temperature measurements are biased toward warmer dust, since it is brighter, so it is likely that colder dust is present deep inside The Brick.

Both an excess of CO in Galactic Center gas, and freezeout during gravitational collapse, may result in a change in the effective equation of state of the gas.
In the dense molecular medium ($n\gtrsim10^3$ \percc) that comprises The Brick, CO is the dominant gas-phase coolant \citep{Ginsburg2016}.
If the CO abundance is \rr{greater} than in the solar neighborhood (\S \ref{sec:wrapupdiscussion}), we expect more efficient cooling in the lower-density outskirts of CMZ clouds.
By contrast, as the clouds collapse to \rr{greater} density, there may be a point at which the CO has frozen out to the point that it is no longer the dominant coolant, but where the densities are still too low for dust to be \rr{an} efficient \rr{coolant}.
We suggest that \rr{systematic} variations in the cooling function \rr{as a function of density or column density} should be explored in future simulations of CMZ cloud thermodynamics like those in \citet{Clark2013}.

\subsection{Broader implications \& future applications}
The prevalence of CO ice in our own Galactic Center hints that ice is likely widespread in galactic centers generally.
At least in the local universe, then, \rr{JWST} observations using the long-wavelength narrowband filters should carefully treat CO absorption in addition to extinction effects.
Our Figure \ref{fig:ncovsav} gives a\rr{n} empirical tool to link extinction and ice, though we caution that the large scatter demonstrated in that plot limits the usefulness of the linear relation given in its legend.

The easy detection of ices in \rr{the F466N} narrow band filter also opens opportunities to better understand dust and ice in the ISM and to understand cloud structures.
While NIRSpec will be capable of studying hundreds of stars with ice absorption in detail, NIRCam observations can easily measure tens of thousands of sightlines simultaneously, enabling detailed correlation analyses like those shown in Fig. \ref{fig:ncovsav}.
By adding comparable-resolution gas observations from ALMA, it should be possible to trace the freezeout of CO from gas to ice in detail.
With a few other bands, such as F300M and F335M, it will be possible to track \water and \methanol ice and determine when and how much CO is incorporated into \water ice, which freezes at a substantially \rr{greater} temperature.

The overlap of the CO ice feature with Pf$\beta$+\hue also opens the possibility of making $\sim0.1\arcsec$ resolution maps of CO ice absorption \rr{against the diffuse ionized emission}.
From the Pa$\alpha$/Br$\alpha$ ratio, we can determine the dust extinction on a per-pixel basis, which will enable specific measurement of CO ice absorption from the \rr{now known} Br$\alpha$/Pf$\beta$ ratio.
Since the CO ice feature affects the Pf$\beta$ line, but CO gas does not, this approach will also allow us to distinguish whether ice or gas is the dominant absorber on most sightlines.

\section{Conclusions}
\label{sec:conclusion}

We report observations of G0.253+0.016, an infrared dark cloud known as ``The Brick,'' with JWST's NIRCam in narrow-band filters.
We produce a crossmatched photometric catalog using the \texttt{crowdsource} package.
We find \nstarstotal unique sources, of which \nstarsallgood are detected in all six photometric bands.

In this first publication on these data, we show that there is significant absorption toward stars in the F466N band, which is caused by CO ice and gas.
We argue that ice is predominant along most sightlines and provide modeling results to show the effect of ice and gas absorption on this and other JWST filters.
While CO absorption is a suitable explanation for the observed F466N absorption, the quantities of both ice and gas required to produce the observed absorption are in some tension with the observed line-of-sight column density.
This result indicates that the standard abundance of CO ($X_{CO}=10^{-4}$) and/or the dust-to-gas ratio (10$^{-2}$) are too low for the Galactic Center environment.

 \textit{Acknowledgements}
 \rr{We thank the referee for a helpful and very detailed report, particularly on their review of labeling conventions and figure clarity.}
 We thank Eddie Schlafly and Andrew Saydjari for their assistance with \texttt{crowdsource} technical issues.
 AG acknowledges support from STSCI grant JWST-GO-02221.001-A, and from the NSF through AST 2008101, AST 220651, and CAREER 2142300.
 CB gratefully  acknowledges  funding  from the National  Science  Foundation  under  Award  Nos. 1816715, 2108938, 2206510, and CAREER 2145689, as well as from the National Aeronautics and Space Administration through the Astrophysics Data Analysis Program under Award No. 21-ADAP21-0179 and through the SOFIA archival research program under Award No.  09$\_$0540. 
 XL\ acknowledges support from the National Natural Science Foundation of China (NSFC) through grant No.\ 12273090, and the Natural Science Foundation of Shanghai (No.\ 23ZR1482100).
 JDH gratefully acknowledges financial support from the Royal Society (University Research Fellowship; URF\textbackslash R1\textbackslash 221620).

\bibliography{bibliography}

\appendix

\section{Comparison to GALACTICNUCLEUS}
\label{appendix:galnuc}
The GALACTICNUCLEUS (GN) \rr{near-infrared} survey \citep{Nogueras-Lara2019,Nogueras-Lara2021} partially overlaps with the targeted field of view \rr{presented here}.
\rr{GN} has \rr{better} resolution than VVV and therefore presents a better match to our data set, but we chose to use VVV as our primary astrometric reference in Section \ref{sec:dataprocessing} because GN covers only about half of the southern field we observed.
The \rr{greater} resolution and sensitivity of GN, however, mean that it is more appropriate for photometric comparison.

\begin{figure*}[!htp]
    \centering
    \includegraphics[width=0.8\textwidth]{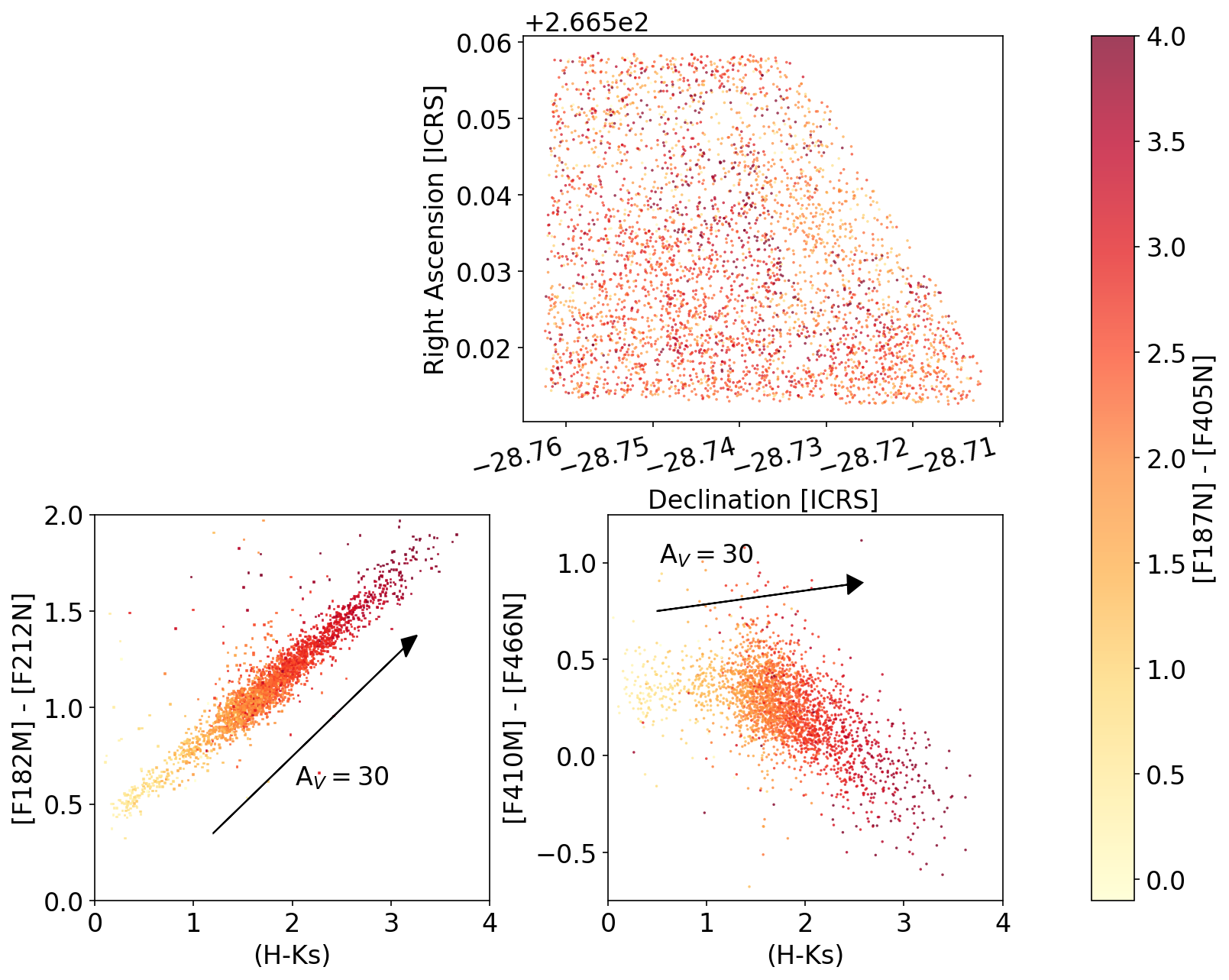}
    \caption{Spatial plot (top) and color-color diagrams (bottom) for the subset of our sample cross-matched with the GALACTICNUCLEUS survey \citep{Nogueras-Lara2021}.
    The left CCD shows our [F182M]-[F212N] color against the GN (H-Ks) color, demonstrating that there is very good correlation between these colors and justifying the use of [F182M]-[F212N] as an extinction estimator.
    The right CCD shows our [F410M]-[F466N] against GN (H-Ks) color, demonstrating that the F466N \rr{deficit} is anticorrelated with (H-Ks) color.
    }
    \label{fig:galnuc_ccd}
\end{figure*}

To verify our use of narrow- and medium-band filters \rr{in our} extinction measurements, we cross-matched our catalog to the GN catalog and produced color-color diagrams using the GN (H-Ks) color as a more typical tracer of extinction.
We found the closest match in our catalog to each GN source and kept all sources with a separation of $<0.2$ \arcsec.
In the field in which GN overlaps our observations, there are a total of 16,021 GN sources and 69,918 sources detected in all three of our short-wavelength bands.
Of the GN sources, 14,557 sources have JWST sources within 0.2 arcseconds, of which 6,844 pass quality criteria specified in \S \ref{sec:catalogxmatch} for all JWST short-wavelength filters and 3,958 pass quality criteria for all six filters.
Figure \ref{fig:galnuc_ccd} shows that we reproduce the same qualitative result as shown in Figure \ref{fig:co_color} with these data.
This plot demonstrates that our color used to measure extinction in the JWST data, [F182M]-[F212N] color, is well-correlated with the standard ground-based (H-Ks) color.

\section{Ice-affected filters}
\begin{figure}
    \includegraphics[]{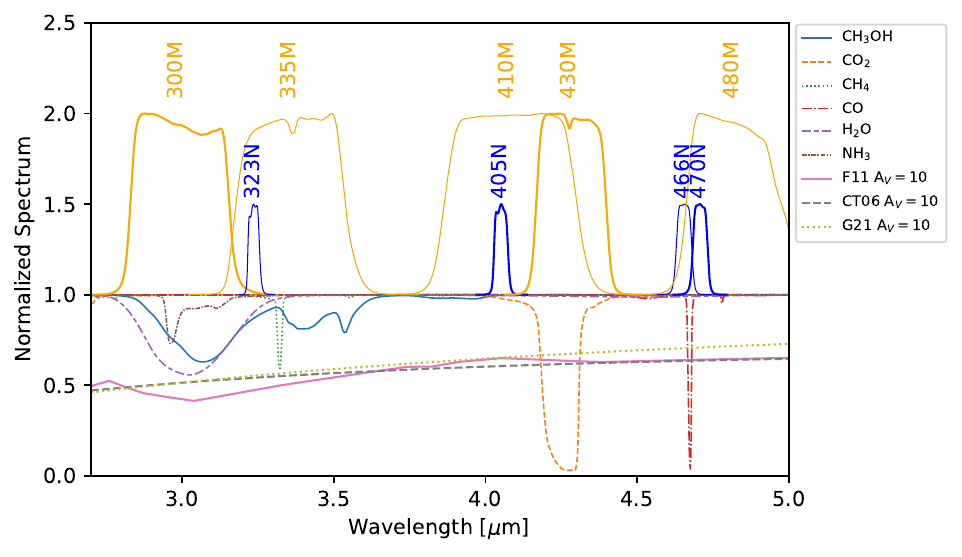}
    \caption{Plot of ice absorption and JWST NIRCam filter profiles.
    The normalized filter profiles are shown as the positive features above 1.0\rr{; all are peak-normalized}.
    Orange colors are medium-band, blue are narrow-band.
    \rr{The blue are scaled to a lower peak to avoid overlapping labels.}
    The ice absorption profiles from several ices are shown in the bottom part of the plot; each ice is plotted with a column density $N(ice)=10^{18}$ \persc.
    For many of these molecules, this is an unrealistically high column density, but it is helpful to illustrate where absorption occurs.
    The \rr{lower panel includes} extinction profiles for three different models: \citet{Chiar2006}, \citet{Fritz2011}, and \citet{Gordon2021}.
    }
    \label{fig:iceaffectedfilters}
\end{figure}

We demonstrate in this paper that ice absorption affects at least the F466N filter in Galactic center photometry.
We highlight the narrow- and medium-band NIRCam filters that are potentially affected by ices in Figure \ref{fig:iceaffectedfilters}.
This plot was made using optical constants from the  JPL Optical Constants Database\footnote{\url{https://ocdb.smce.nasa.gov/page/toc}} using the \texttt{icemodels} package\footnote{\url{https://github.com/keflavich/icemodels/}}.
\rr{This figure provides a quick-look tool for determining how likely an observation in a given filter is to be affected by ice absorption, and therefore is a good first stop to check if one encounters unexpected colors in the long-wavelength NIRCam bands.}

\end{document}